\documentclass[twocolumn,pra,superscriptaddress,showpacs]{revtex4}
\usepackage[utf8]{inputenc}
\usepackage{amsmath}
\usepackage{amssymb}
\usepackage{graphicx}
\usepackage{esint}
\usepackage[unicode=true,pdfusetitle,
 bookmarks=true,bookmarksnumbered=false,bookmarksopen=false,
 breaklinks=false,pdfborder={0 0 1},backref=section,colorlinks=false]
 {hyperref}
\hypersetup{
 colorlinks,citecolor=blue}
\usepackage{breakurl}

\makeatletter
\@ifundefined{textcolor}{}
{%
 \definecolor{BLACK}{gray}{0}
 \definecolor{WHITE}{gray}{1}
 \definecolor{RED}{rgb}{1,0,0}
 \definecolor{GREEN}{rgb}{0,1,0}
 \definecolor{BLUE}{rgb}{0,0,1}
 \definecolor{CYAN}{cmyk}{1,0,0,0}
 \definecolor{MAGENTA}{cmyk}{0,1,0,0}
 \definecolor{YELLOW}{cmyk}{0,0,1,0}
}

\usepackage{subfigure}\usepackage{epsfig}\usepackage{amsfonts}\usepackage{mathrsfs}\usepackage{CJK}\usepackage[toc,page,title,titletoc,header]{appendix}

\makeatother

\begin{document}

\title{Suppression of two-body collisional loss in an ultracold gas via the
Fano effect}

\author{Jianwen Jie}

\affiliation{Department of Physics, Renmin University of China, Beijing, 100872,
China}

\author{Yawen Zhang}

\affiliation{Department of Physics, Renmin University of China, Beijing, 100872,
China}

\author{Peng Zhang}

\affiliation{Department of Physics, Renmin University of China, Beijing, 100872,
China}

\affiliation{Beijing Computational Science Research Center, Beijing, 100084, China}

\affiliation{Beijing Key Laboratory of Opto-electronic Functional Materials \&
Micro-nano Devices (Renmin University of China)}

\begin{abstract}
The Fano effect (U. Fano, Phys. Rev. \textbf{15},
1866 (1961))
shows that an inelastic scattering process can be suppressed when
the output channel (OC) is coupled to an isolated bound state. In this paper
we investigate the application of this effect for the suppression of two-body collisional losses
of ultracold atoms. The Fano effect is originally
derived via a first-order perturbation treatment for coupling
between the incident channel (IC) and the OC. We generalize the Fano effect to systems with
arbitrarily strong IC--OC couplings. We analytically prove that, in a system with one IC and one OC,
when the inter-atomic interaction potentials are real functions of the inter-atomic distance,
the exact $s$-wave inelastic scattering
amplitude can always be suppressed to \emph{zero} by coupling between the IC or the OC (or both of them) and an extra isolated bound state.
We further show that when the low-energy inelastic collision between two ultracold atoms is suppressed by this effect, the real part of the elastic scattering length between the atoms is still possible to be much larger than the range of inter-atomic interaction.
In addition, when open
scattering channels are coupled to two bound states, with the help
of the Fano effect, independent control
of the elastic and inelastic scattering amplitudes of two ultracold
atoms can be achieved. Possible experimental realizations of our scheme are also discussed.
\end{abstract}

\maketitle

\section{introduction}

In ultracold gases of neutral atoms prepared in excited internal
states (e.g., excited hyperfine states corresponding to the electronic
ground level of alkali atoms or long-lived excited states of alkali-earth
(like) atoms), two-body collisional losses can be induced by inelastic
scattering processes \cite{pethick,rb,ca,yb2,Rb87loss,Rb87loss2,Csloss,Rb85loss3,Rb85loss,Rb85loss2,
Li6loss,Li6loss2,NJP,molecule,rb2,paul,sem}.
In these processes the atoms
can jump to the lower internal states and gain a large amount of kinetic energy.
Two-body collisional losses can shorten the lifetime of the atomic
gases. For instance, in the Bose--Einstein condensate of $^{{\rm 87}}$Rb
atoms in the hyperfine state $|F=2,m_{F}=1\rangle$ and the ultracold gas
$^{173}$Yb atoms in the $^{3}$P$_{0}$ states,
two-body collisional loss rates are of the order of $10^{-13}{\rm cm}^{3}/{\rm s}$
\cite{rb,Rb87loss} and $10^{-11}{\rm cm}^{3}/{\rm s}$ \cite{yb2}, respectively.
This means that for ultracold gases of these atoms having typical densities of
$10^{14}/{\rm cm}^{3}$, the lifetime can be reduced to much less
than 1 s or even less than 1 ms. In most
experiments of optically trapped ultracold gases, the atoms are prepared
in the lowest internal states so that two-body inelastic scattering
can be avoided.

Nevertheless, a lot of interesting physics can be studied with ultracold
gases of atoms prepared in excited internal states. For instance,
the physics of spin-2 Bose Einstein condensation can be studied with ultracold $^{87}$Rb
atoms with $F=2$ \cite{F2BEC}. Physics
related to spin-exchange processes and the Kondo effect can be studied
with a mixture of ultracold alkali-earth (like) atoms in the ground
$^{1}$S$_{0}$ and excited $^{3}$P$_{0}$ states \cite{sr,yb1,yb2}.
To obtain such ultracold gases with sufficiently long lifetimes, it is important
to study how to suppress the two-body inelastic scattering processes
between ultracold atoms \cite{Rb85loss2,molecule,rb2,sem,NJP,Rb85loss3}.

In 1961, Ugo Fano found that inelastic scattering can be significantly
suppressed if the output channel (OC) of that process is coupled to
an isolated bound state \cite{fano}. The Fano effect can be understood
as the result of destructive interference between the quantum transition
from the incident channel (IC) to the OC and the transition from the isolated bound state to the OC. In the original derivation of the Fano effect, coupling between the IC
and OC is treated as a first-order perturbation \cite{fano,fano2}.
This perturbative treatment has also been used in the previous study for the application of the Fano effect on the suppression of collisional loss in ultracold gases  \cite{sem}.


In this paper, we go beyond this first-order perturbation approximation
and investigate the Fano effect in the two-atom scattering problem
with arbitrarily strong
IC--OC coupling. Then, we study the application of this effect
for the suppression of the two-body collisional losses in ultracold
gases. The main results and the structure of this paper can be summarized
as follows:

In Sec. II we study the Fano effect in a two-atom scattering problem
in three-dimensional space, with one IC and one OC. Here each
channel corresponds to a two-atom internal state. With an analytical
calculation for the exact inelastic scattering amplitude, we prove
that when the inter-atomic interaction potentials are real functions
of the distance between these two atoms, the $s$-wave inelastic
scattering amplitude can always be suppressed to \emph{zero }when
an isolated bound state is coupled to the IC or the OC, or both of
them (Fig. 1). Our result is applicable for systems with arbitrary
IC--OC coupling intensities and incident kinetic energy. In particular,
we prove that this suppression effect can occur even when the bound
state is only coupled to the IC and not directly coupled to the OC, as shown in Fig. 1(c). We show
that the suppression effect can be understood as resulting from
destructive interference between the direct transition from the IC to the OC, and
an indirect transition from the IC to the isolated bound state and then back
to the IC, and then to the OC.

\begin{figure}
\includegraphics[width=8.5cm]{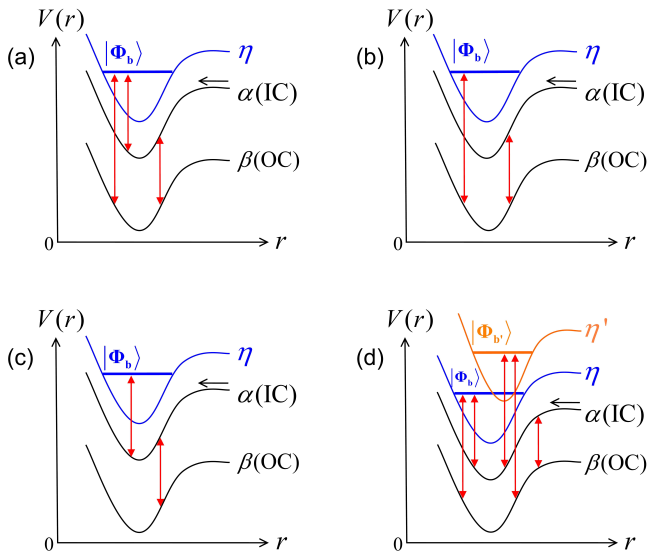}
\caption{(color online) Multi-channel models studied in Sec. II (a--c) and Sec. IV (d).
Here $r$ is the distance between the two atoms.
The black curves represent the interaction potentials
for the IC $\alpha$ and OC $\beta$ of the
inelastic scattering processes, while the
blue and orange curves represent the interaction potentials
for the closed channels $\eta$ and $\eta^\prime$, respectively.
The red arrows represent inter-channel coupling.
In Sec. II, we consider cases where a bound state $|\Phi_b\rangle$ in the closed
channel $\eta$ is coupled to both the IC $\alpha$ and OC $\beta$ (a),
only coupled to
the OC $\beta$
(b), or only coupled to the IC $\alpha$ (c). In Sec. IV, we consider the system where the open channels $\alpha$
and $\beta$ are coupled to two isolated bound states$|\Phi_b\rangle$ and $|\Phi_{b^\prime}\rangle$ in the
$\eta$ and $\eta\prime$ channels (d).}
\end{figure}

Our results imply that two-body collisional loss in an ultracold
gas may be completely suppressed when there is one OC in the inelastic
scattering processes. In Sec. III we further show that when this loss
is suppressed by the Fano effect, it is still possible for the elastic scattering length between
the two ultracold atoms to be either large
or small. That is, it is possible to obtain ultracold gases of atoms
in excited internal states with strong inter-atom interactions
and negligible collisional loss rates. To our knowledge, this result
has not been obtained in previous studies on the control of two-body
collisional loss in ultracold gases. In this section we also discuss
the possible experimental realizations of our approach.

In Sec. IV we investigate the independent control of elastic and inelastic
collisions between ultracold atoms. We study the system where the IC
and OC are coupled to two bound states (Fig. 1(d)). We show that,
for this system, when the inelastic scattering amplitude is suppressed
to zero by the Fano effect, the elastic scattering
length in the IC can be tuned to any value by altering the energies of these two
bound states.

These results are helpful for the study of ultracold gases of
atoms prepared in excited internal states. Moreover, our generalization
of the Fano effect for systems with strong IC--OC coupling is also
useful for the study of inelastic scattering processes in other physical
systems.

\section{Fano effect in the system with strong IC--OC coupling}

We consider the three-channel scattering problem of two ultracold
atoms shown in Fig.~1(a--c). The three channels $\alpha$, $\beta$,
and $\eta$ correspond to two-atom internal states $|\alpha\rangle_{I}$,
$|\beta\rangle_{I}$, and $|\eta\rangle_{I}$, respectively. In
natural units $\hbar=m=1$, with $m$ the single-atom mass, we can express
the Hamiltonian of our system as
\begin{equation}
H={\bf p}^{2}+\sum_{j=\alpha,\beta,\eta}E_{j}|j\rangle_{I}\langle j|+V(r),\label{hab}
\end{equation}
where ${\bf p}$ and ${\bf r}$ are the relative momentum and relative
coordinate of the two atoms, respectively, and $r=|{\bf r}|$. The
energy $E_{j}$ ($j=\alpha,\beta,\eta$) is the threshold energy of
channel $j$, with $E_{\eta}>E_{\alpha}>E_{\beta}$. In Eq. (\ref{hab}),
$V(r)$ is the interaction potential of the two atoms, and is given
by
\begin{equation}
V(r)=\sum_{l,j=\alpha,\beta,\eta}V_{lj}(r)|l\rangle_{I}\langle j|.\label{vr}
\end{equation}
Here $V_{jj}(r)$ ($j=\alpha,\beta,\eta$) is the potential of channel
$j$, while $V_{lj}(r)=V_{jl}(r)$ ($l\neq j$) is the inter-channel
coupling. For the systems shown in Fig. 1(b) and Fig. 1(c),
we have $V_{\alpha\eta}=0$ and $V_{\beta\eta}=0$, respectively.
In this paper we consider systems where all the components
of $V$ are a real function of $r$.

We consider the case where the two atoms are incident from channel
$\alpha$, and the incident state is near resonant to an isolated
$s$-wave bound state $|\Phi_b\rangle$ in channel $\eta$. In this case, channels $\alpha$
and $\beta$ are the IC and OC of the inelastic scattering process,
respectively. The $s$-wave inelastic scattering amplitude $f_{\beta\alpha}$
from channel $\alpha$ to $\beta$ can be expressed as a function
of the scattering energy $E_{{\rm s}}$ and the energy $\epsilon_{b}$
of $|\Phi_b\rangle$. In the following we will prove
that $f_{\beta\alpha}$ can always be suppressed to \emph{zero} by
coupling between the bound
state in channel $\eta$ and the channels $\alpha$ and/or $\beta$, no matter how strong the IC--OC
coupling $V_{\alpha\beta}$. That is, for any given value of $E_{{\rm s}}$,
there always exists a real energy $\tilde{\epsilon}_{b}$, which leads
to $f_{\beta\alpha}(E_{{\rm s}},\epsilon_{b}=\tilde{\epsilon}_{b})=0$.

In the following subsections we will first derive the analytical
expression of $f_{\beta\alpha}$, and then calculate the non-diagonal
element of the $K$-matrix for our system. This element is proportional
to $f_{\beta\alpha}$ and easier to study. We will prove our result
by analyzing the character of this $K$-matrix element.

\subsection{Scattering amplitude}

In this subsection we calculate the $s$-wave scattering amplitude
with the method in Ref. \cite{rmp}. In our system the Hilbert space
$\mathcal{H}$ can be expressed as $\mathcal{H}=\mathcal{H}_{R}\otimes\mathcal{H}_{I}$,
with $\mathcal{H}_{R}$ being the Hilbert space for the inter-atomic relative
motion in the spatial space and $\mathcal{H}_{I}$ representing the two-atom internal state. We use $|\rangle$ to denote the state in $\mathcal{H}$, $|\rangle_{R}$\ for the state
in $\mathcal{H}_{R}$, and\ $|\rangle_{I}$\ for the state in ${\cal H}_{I}$.
The scattering amplitude from channel $l$ to channel
$j$ ($l$, $j=\alpha,\beta$) is defined as
\begin{equation}
f_{jl}=-2\pi^{2}\langle\Psi_{k_{j},j}^{(0)}|V|\Psi_{k_{l},l}^{(+)}\rangle,\label{f}
\end{equation}
where $|\Psi_{k_{l},l}^{(+)}\rangle$ is the $s$-wave component of
the out-going scattering state with respect to the incident momentum ${\bf k}_{l}$
and incident channel $l$, and the state $|\Psi_{k_{j},j}^{(0)}\rangle$
is defined as $|\Psi_{k_{j},j}^{(0)}\rangle=|\psi_{k_{j},j}^{(0)}\rangle_{R}|j\rangle_{I}$,
with $|\psi_{k_{j},j}^{(0)}\rangle_{R}$ the $s$-wave component of
the eigen-state $|{\bf k}_{j}\rangle_{R}$ of the relative momentum
operator ${\bf p}$. Here we have $k_{l(j)}=|{\bf k}_{l(j)}|$. Notice that the $s$-wave
states $|\Psi_{k_{l},l}^{(+)}\rangle$ and $|\Psi_{k_{j},j}^{(0)}\rangle$
are independent of the directions of the momentum ${\bf k}_{l}$
and ${\bf k}_{j}$, respectively. Due to energy conservation,
the momentum $k_{l,j}$ satisfies
\begin{equation}
k_{l}^{2}+E_{l}=k_{j}^{2}+E_{j}\equiv E_{{\rm s}},\label{se}
\end{equation}
where $E_{{\rm s}}$ is defined as the scattering energy.

We can obtain the scattering amplitude by solving the Lippman--Schwinger
equation satisfied by the scattering state $|\Psi_{k_{l},l}^{(+)}\rangle$.
This equation can be expressed as (Ref. \cite{rmp}, Appendix A)
\begin{equation}
|\Psi_{k_{l},l}^{(+)}\rangle=|\Psi_{k_{l},l}^{(\alpha\beta+)}\rangle+G^{(\alpha\beta)}(E_{{\rm s}})W|\Psi_{k_{l},l}^{(+)}\rangle,\label{psis}
\end{equation}
where the operator $W$ is defined as
\begin{equation}
W=V_{\eta\alpha}(r)|\eta\rangle_{I}\langle\alpha|+V_{\eta\beta}(r)|\eta\rangle_{I}\langle\beta|+h.c.,\label{w-1-1}
\end{equation}
and describes the coupling between channels $\alpha,\beta$, and $\eta$.
Here $|\Psi_{k_{l},l}^{(\alpha\beta+)}\rangle$ is the $s$-wave component
of the out-going scattering state for the case with $W=0$, with respect
to the incident channel $l$ and incident momentum ${\bf k}_{l}$, and
$G^{(\alpha\beta)}(E)$ is the Green's operator for this case. It
is given by
\begin{equation}
G^{(\alpha\beta)}(E)=\frac{1}{E+i0^{+}-(H-W)}.\label{gab}
\end{equation}

As shown above, we consider the case where $E_{{\rm s}}$ is near
resonant to an isolated $s$-wave bound state $|\Phi_{b}\rangle\equiv|\phi_{b}\rangle_{R}|\eta\rangle_{I}$
in channel $\eta$. Here $|\phi_{b}\rangle_{I}$ satisfies the eigen-equation
\begin{equation}
H_{\eta}|\phi_{b}\rangle_{R}\equiv\left[{\bf p}^{2}+V_{\eta\eta}(r)+E_{\eta}\right]|\phi_{b}\rangle_{R}=\epsilon_{b}|\phi_{b}\rangle_{R}\label{ee-1}
\end{equation}
of the self-Hamiltonian $H_{\eta}$ of channel $\eta$, and ``near
resonant'' means that $E_{{\rm s}}$ is close to $\epsilon_{b}$. In this case, we can neglect the
contribution from other eigen-states of $H_{\eta}$. Under this single-resonance
approximation, the Green's operator $G^{(\alpha\beta)}(E)$ can be
re-expressed as
\begin{equation}
G^{(\alpha\beta)}(E)=\frac{1}{E+i0^{+}-h}+\frac{|\Phi_{b}\rangle\langle\Phi_{b}|}{E-\epsilon_{b}},\label{gg2}
\end{equation}
where
\begin{equation}
h={\bf p}^{2}+\sum_{j=\alpha,\beta}E_{j}|j\rangle_{I}\langle j|+\sum_{l,j=\alpha,\beta}V_{lj}(r)|l\rangle_{I}\langle j|\label{sh}
\end{equation}
is the ``self-Hamiltonian'' of channels $\alpha$ and $\beta$.
With Eq. (\ref{gg2}), we can analytically solve the Lippman--Schwinger
equation (\ref{psis}) for the scattering state $|\Psi_{k_{l},l}^{(+)}\rangle$,
and thus obtain the $s$-wave scattering amplitude $f_{jl}(E_{{\rm s}})$
defined in Eq. (\ref{f}) (Ref. \cite{rmp}, Appendix B):
\begin{equation}
f_{jl}(E_{{\rm s}},\epsilon_{b})=f_{jl}^{(\alpha\beta)}(E_{{\rm s}})-2\pi^{2}\frac{A_{j}(E_{{\rm s}})A_{l}(E_{{\rm s}})}{B(E_{{\rm s}})-\epsilon_{b}},\label{fji}
\end{equation}
where $f_{jl}^{(\alpha\beta)}(E_{{\rm s}})$ is the scattering amplitude
for the case with $W=0$, and the functions $A_{l(j)}(E_{{\rm s}})$
and $B(E_{{\rm s}})$ are defined as
\begin{eqnarray}
A_{l(j)}(E_{{\rm s}}) & = & \langle\Phi_{b}|W|\Psi_{k_{l(j)},l(j)}^{(\alpha\beta+)}\rangle;\label{aji}\\
B(E_{{\rm s}}) & = & E_{{\rm s}}-\langle\Phi_{b}|WG^{(\alpha\beta)}(E_{{\rm s}})W|\Phi_{b}\rangle.\label{c}
\end{eqnarray}

\subsection{$S$-matrix and $K$-matrix}

In this subsection we introduce the $S$-matrix and $K$-matrix related
to the $s$-wave scattering in our system. In the $s$-wave subspace
the $S$-matrix is a $2\times2$ matrix
\begin{equation}
S(E_{{\rm s}},\epsilon_{b})=\left[\begin{array}{cc}
S_{\alpha\alpha}(E_{{\rm s}},\epsilon_{b}), & S_{\alpha\beta}(E_{{\rm s}},\epsilon_{b})\\
S_{\beta\alpha}(E_{{\rm s}},\epsilon_{b}), & S_{\beta\beta}(E_{{\rm s}},\epsilon_{b})
\end{array}\right].\label{s}
\end{equation}
Here the matrix element $S_{jl}(E_{{\rm s}},\epsilon_{b})$ ($l,j=\alpha,\beta$)
is related to the scattering amplitude via the relation
\begin{equation}
f_{jl}(E_{{\rm s}},\epsilon_{b})=\frac{S_{jl}(E_{{\rm s}},\epsilon_{b})-\delta_{jl}}{2i\sqrt{k_{l}k_{j}}}.\label{fs}
\end{equation}
In Appendix C we show the relation between this $S$-matrix and the
$S$-operator of our system \cite{st}, and prove that this $S$-matrix is a unitary matrix \cite{st}.

In our system the $K$-matrix is defined as \cite{kmatrix}
\begin{eqnarray}
K(E_{{\rm s}},\epsilon_{b}) & = & i\frac{1-S(E_{{\rm s}},\epsilon_{b})}{1+S(E_{{\rm s}},\epsilon_{b})}\label{k}\\
 & \equiv & \left[\begin{array}{cc}
K_{\alpha\alpha}(E_{{\rm s}},\epsilon_{b}), & K_{\alpha\beta}(E_{{\rm s}},\epsilon_{b})\\
K_{\beta\alpha}(E_{{\rm s}},\epsilon_{b}), & K_{\beta\beta}(E_{{\rm s}},\epsilon_{b})
\end{array}\right].\nonumber
\end{eqnarray}
According to this definition, the non-diagonal elements of the $K$-matrix
and $S$-matrix satisfy the relation
\begin{equation}
K_{\beta\alpha}=\frac{-2iS_{\beta\alpha}}{1+{\rm Det}[S]+S_{\alpha\alpha}+S_{\beta\beta}}.\label{ks}
\end{equation}
With direct calculation based on Eqs. (\ref{fji}, \ref{fs}) and
(\ref{ks}), we can obtain the expression of $K_{\beta\alpha}(E_{{\rm s}},\epsilon_{b})$.
Since the terms $S_{jl}$ ($j,l=\alpha,\beta$) and ${\rm Det}[S]$ in Eq. (\ref{ks})
are linear and quadratic functions of the scattering amplitude $f_{ij}$ given by Eq. (\ref{fji}), respectively, $K_{\beta\alpha}(E_{{\rm s}},\epsilon_{b})$ can be expressed as
\begin{equation}
K_{\beta\alpha}(E_{{\rm s}},\epsilon_{b})=\frac{F_{1}(E_{{\rm s}})\epsilon_{b}^2+C_{1}(E_{{\rm s}})\epsilon_{b}+D{}_{1}(E_{{\rm s}})}{F_{2}(E_{{\rm s}})\epsilon_{b}^2+C_{2}(E_{{\rm s}})\epsilon_{b}+D_{2}(E_{{\rm s}})},\label{kab22}
\end{equation}
and we can obtain the coefficients $C_{1,2}$, $D_{1,2}$ and $F_{1,2}$ via
substituting Eqs. (\ref{fji}, \ref{fs}) into Eq. (\ref{ks}). With direct calculation, we are surprised to find that the coefficients $F_{1}(E_{\rm s})$ and $F_{2}(E_{\rm s})$ of the $\epsilon_{b}^2$-terms in Eq. (\ref{kab22}) are exactly zero, i.e., $F_{1}(E_{\rm s})=F_{2}(E_{\rm s})=0$. As a result, $K_{\beta\alpha}(E_{{\rm s}},\epsilon_{b})$ has a simple expression
\begin{equation}
K_{\beta\alpha}(E_{{\rm s}},\epsilon_{b})=\frac{C_{1}(E_{{\rm s}})\epsilon_{b}+D{}_{1}(E_{{\rm s}})}{C_{2}(E_{{\rm s}})\epsilon_{b}+D_{2}(E_{{\rm s}})},\label{kab}
\end{equation}
where the coefficients $C{}_{1,2}(E_{{\rm s}})$ and $D_{1,2}(E_{{\rm s}})$
are given by\begin{widetext}
\begin{eqnarray}
C_{1}(E_{{\rm s}}) & = & 2is_{\beta\alpha}(E_{{\rm s}});\label{c1}\\
C_{2}(E_{{\rm s}}) & = & -1-s_{\alpha\alpha}(E_{{\rm s}})s_{\beta\beta}(E_{{\rm s}})+s_{\alpha\beta}(E_{{\rm s}})s_{\beta\alpha}(E_{{\rm s}})-s_{\alpha\alpha}(E_{{\rm s}})-s_{\beta\beta}(E_{{\rm s}});\label{c2}\\
D{}_{1}(E_{{\rm s}}) & = & -2i\left[s_{\alpha\beta}(E_{{\rm s}})B(E_{{\rm s}})+{\cal A}_{\alpha\beta}(E_{{\rm s}})\right];\label{d1}\\
D{}_{2}(E_{{\rm s}}) & = & -C_{2}(E_{{\rm s}})B(E_{{\rm s}})+s_{\alpha\alpha}(E_{{\rm s}}){\cal A}_{\beta\beta}(E_{{\rm s}})+s_{\beta\beta}(E_{{\rm s}}){\cal A}_{\alpha\alpha}(E_{{\rm s}})-{\cal A}(E_{{\rm s}})\left[s_{\beta\alpha}(E_{{\rm s}})+s_{\alpha\beta}(E_{{\rm s}})\right]\nonumber \\
 &  & +{\cal A}_{\alpha\alpha}(E_{{\rm s}})+{\cal A}_{\beta\beta}(E_{{\rm s}}),\label{d2}
\end{eqnarray}
\end{widetext} with ${\cal A}_{lj}(E_{{\rm s}})=-4\pi^{2}i\sqrt{k_{j}k_{l}}A_{l}(E_{{\rm s}})A_{j}(E_{{\rm s}})$
($l,j=\alpha,\beta$). Here, the functions $A_{l(j)}(E_{{\rm s}})$
and $B(E_{{\rm s}})$ are defined in Eqs. (\ref{aji}) and (\ref{c}),
and $s_{lj}(E_{{\rm s}})$ is the element of the $S$-matrix for the
case with $W=0$.

\subsection{Suppression of inelastic scattering}

Based on our above results, now we prove the central result of this section.

Because the interaction potential in our system is real, the $S$-matrix $S(E_{{\rm s}},\epsilon_{b})$
is a symmetric unitary matrix (Ref. \cite{rmp}, appendix C), and
thus can be formally expressed as
\begin{equation}
S(E_{{\rm s}},\epsilon_{b})=\left(\begin{array}{cc}
\zeta e^{i\xi} & \sqrt{1-\zeta^{2}}e^{i\xi'}\\
\sqrt{1-\zeta^{2}}e^{i\xi'} & -\zeta e^{-i\xi}e^{2i\xi'}
\end{array}\right),\label{s2}
\end{equation}
where $\zeta,$ $\xi$, and $\xi'$ are real numbers and $0<\zeta\leq1$.
Substituting Eq. (\ref{s2}) into Eq. (\ref{ks}), we find that the
non-diagonal $K$-matrix element $K_{\beta\alpha}(E_{{\rm s}},\epsilon_{b})$
can be re-expressed as
\begin{equation}
K_{\beta\alpha}(E_{{\rm s}},\epsilon_{b})=\frac{\sqrt{1-\zeta^{2}}}{-\sin\xi'+\zeta\sin(\xi-\xi')},\label{k2}
\end{equation}
and thus must be real for all values of $E_{{\rm s}}$ and $\epsilon_{b}$.
Using this result and the expression (\ref{kab}) for $K_{\beta\alpha}(E_{{\rm s}},\epsilon_{b})$,
it can be proved that (Appendix D) the ratio $D_{1}(E_{{\rm s}})/C_{1}(E_{{\rm s}})$
is always real. Thus, according to Eq. (\ref{kab}), non-diagonal
element $K_{\beta\alpha}(E_{{\rm s}},\epsilon_{b})$ of the $K$-matrix
becomes zero when the energy $\epsilon_{b}$ of the bound state in
channel $\eta$ takes the value
\begin{equation}
\epsilon_{b}=-\frac{D_{1}(E_{{\rm s}})}{C_{1}(E_{{\rm s}})}.\label{cc}
\end{equation}
Furthermore, according to Eqs. (\ref{ks}) and (\ref{fs}), we have
\begin{equation}
K_{\beta\alpha}(E_{{\rm s}},\epsilon_{b})\propto S_{\beta\alpha}(E_{{\rm s}},\epsilon_{b})\propto f_{\beta\alpha}(E_{{\rm s}},\epsilon_{b}).\label{prop}
\end{equation}
Therefore, under the condition in Eq. (\ref{cc}), we have
\begin{eqnarray}
f_{\beta\alpha} & = & 0,
\end{eqnarray}
i.e., the inelastic scattering from the IC $\alpha$ to the OC $\beta$
is completely suppressed by coupling $W$ between these two channels
and the bound state $|\Phi_{b}\rangle$ in the closed channel $\eta.$
Because we do not treat the IC--OC coupling $V_{\alpha\beta}$
as a perturbation in our proof, our result is applicable to systems with arbitrarily
strong IC--OC coupling.

Our proof shows that the inelastic scattering amplitude can
be suppressed as long as the inter-channel coupling $W$ defined in Eq.
(\ref{w-1-1}) is nonzero. This is regardless of whether the coupling $V_{\beta\eta}$
between the bound state $|\Phi_{b}\rangle$ and the OC $\beta$ is
zero or nonzero. When $V_{\beta\eta}\neq0$, the suppression effect
can be understood as a result of interference between the quantum
transition from channel $\alpha$ to channel $\beta$ and the one
from $|\Phi_{b}\rangle$ to channel $\beta$. Nevertheless, in
systems with $V_{\beta\eta}=0$ and $V_{\alpha\eta}\neq0$, i.e., the
system shown in Fig. 1(c), this effect
is not attributable to direct interference of these quantum transitions.

\begin{figure*}
\includegraphics[width=17cm]{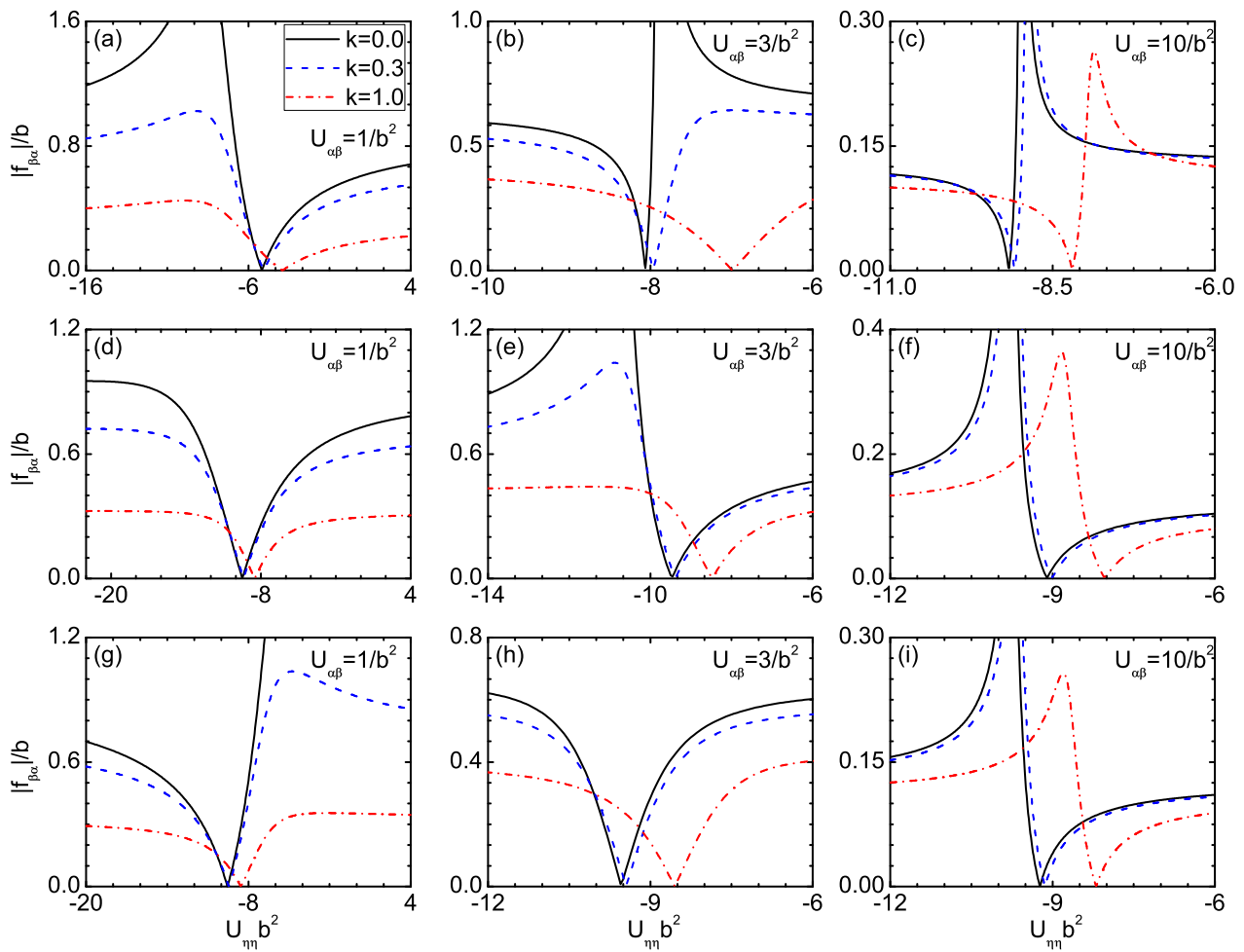}
\caption{(color online) The absolute value $|f_{\beta\alpha}|$ of the inelastic
scattering amplitude in the square-well model, as a function
of the potential energy $U_{\eta\eta}$ of channel $\eta$.
In (a--c), we show
results for cases where $U_{\alpha\eta}=2/b^{2}$, $U_{\beta\eta}=3/b^{2}$, i.e., the cases where the closed channel $\eta$ is coupled to both IC
$\alpha$ and OC $\beta$ (the cases in Fig. 1(a)). In (d--f),
we show results for cases where $U_{\alpha\eta}=0$,
$U_{\beta\eta}=3/b^{2}$, i.e., the cases where the closed channel $\eta$ is only coupled
to OC $\beta$ (the cases in Fig. 1(b)). In (g--i), we show results for cases where $U_{\alpha\eta}=3/b^{2}$, $U_{\beta\eta}=0$,
i.e., the cases where the closed channel $\eta$ is only coupled to the IC $\alpha$ (the cases in Fig. 1(c)).
Here we consider
systems with potential energies $U_{\alpha\alpha}=-1/b^{2}$, $U_{\beta\beta}=-2.5/b^{2}$;
threshold energies $E_{\alpha}=0$, $E_{\beta}=-0.5/b^{2}$; inter-channel
coupling $U_{\alpha\beta}=1/b^{2}$ (a,d,g), $3/b^{2}$ (b,e,h) and
$10/b^{2}$ (c,f,i); and incident momentum $k_{\alpha}=0$ (solid black line),
$0.3/b$ (dashed blue line), and $1/b$ (dash-dotted red line).}
\end{figure*}

To understand the suppression effect in this special case, we consider
a system where the IC--OC coupling $V_{\alpha\beta}$ is very weak
and can be treated as a first-order perturbation. For this system
the inelastic scattering amplitude $f_{\beta\alpha}$ can be approximated
as $f_{\beta\alpha}\approx-2\pi^{2}\int drr^{2}\psi_{\beta}^{\ast}(r)V_{\alpha\beta}(r)\psi_{\alpha}(r)$,
where $\psi_{\alpha(\beta)}(r)$ is the component of the $s$-wave
scattering wave function in channel $\alpha$ ($\beta$) for the case
with $V_{\alpha\beta}=0$. In the presence of the coupling between the
 IC $\alpha$ and the
bound state $|\Phi_{b}\rangle$ in channel $\eta$, the wave function
$\psi_{\alpha}(r)$ can be formally
expressed as $\psi_{\alpha}(r)=\psi_{\alpha}^{({\rm bg})}(r)+\delta\psi_{\alpha}(\epsilon_{b},r)$.
Here $\psi_{\alpha}^{({\rm bg})}(r)$ is the $s$-wave scattering
wave function in channel $\alpha$ for the case with $V_{\alpha\eta}=V_{\alpha\beta}=0$,
and the $\epsilon_{b}$-dependent wave function $\delta\psi_{\alpha}(\epsilon_{b},r)$
is the modification induced by $V_{\alpha\eta}$. This term describes
the change of the atomic wave function in channel $\alpha$, which
is induced by the second-order process where the atoms transit from
channel $\alpha$ to $|\Phi_{b}\rangle$ and then return to $\alpha$.
Accordingly, the inelastic scattering
amplitude $f_{\beta\alpha}$ can be expressed
as
\begin{eqnarray}
f_{\beta\alpha}(E_{{\rm s}},\epsilon_{b}) & \approx & -2\pi^{2}\int drr^{2}\psi_{\beta}^{\ast}(r)V_{\alpha\beta}(r)\psi_{\alpha}^{({\rm bg})}(r)\nonumber \\
 &  & -2\pi^{2}\int drr^{2}\psi_{\beta}^{\ast}(r)V_{\alpha\beta}(r)\delta\psi_{\alpha}(\epsilon_{b},r).\nonumber \\
\label{f1p}
\end{eqnarray}
Eq. (\ref{f1p}) clearly shows that the inelastic scattering amplitude
includes contributions from the transition processes from the states $\psi_{\alpha}^{({\rm bg})}(r)$
and $\delta\psi_{\alpha}(\epsilon_{b},r)$ to
the state $\psi_{\beta}(r)$. When the interference
of these two transition processes is destructive, the inelastic scattering
can be suppressed. This analysis shows that, in a system where the
bound state $|\Phi_{b}\rangle$ is only coupled to IC $\alpha$
and not coupled to OC $\beta$, the suppression of the inelastic
scattering can be understood as a result of destructive interference
between the direct transition from channel $\alpha$ to $\beta$ and the
indirect transition process along the path $\alpha\rightarrow|\Phi_{b}\rangle\rightarrow\alpha\rightarrow\beta$.

\subsection{Illustration}

Now we illustrate our results with a simple multi-channel square-well
model. In this model the potential $V_{lj}(r)$ ($l,j=\alpha,\beta,\eta$)
defined in Eq. (\ref{vr}) is given by
\begin{equation}
V_{lj}(r)=\left\{ \begin{array}{c}
U_{lj},\ {\rm for}\ r<b\\
0,\ {\rm for}\ r>b
\end{array},\right.\label{vsw}
\end{equation}
with $b$ the range of these potentials. We further choose the threshold
energies $E_{\alpha,\beta,\eta}$ to satisfy
\begin{equation}
E_{\eta}=\infty,\ E_{\alpha}=0,\ E_{\beta}<0.\label{ej}
\end{equation}
We calculate the inelastic scattering amplitude $f_{\beta\alpha}$
from the higher channel $\alpha$ to the lower channel $\beta$, for
cases where the incident state is near resonant to the lowest bound
state in channel $\eta$.
Figure 2 shows $|f_{\beta\alpha}|$ as
a function of the potential energy $U_{\eta\eta}$ of channel $\eta$.
Here, we consider cases where the closed channel $\eta$ is coupled to both the IC $\alpha$ and OC
$\beta$ (Fig. 2(a-c)), and cases where channel $\eta$ is only coupled to the IC $\alpha$ (Fig. 2(d-f))
or OC $\beta$ (Fig. 2(g-i)).
It is clearly shown that in all of these cases, for the system with any incident momentum
$k_{\alpha}$ and IC--OC coupling $U_{\alpha\beta}$, the inelastic
amplitude $f_{\beta\alpha}$ can always be suppressed to zero.

\section{suppression of inelastic scattering processes in ultracold gases}

In this and the next section, we study the application of our results in
ultracold gases. In this system, if there is only one possible two-atom
inelastic scattering process (e.g., the scattering from channel $\alpha$
to channel $\beta$) then the two-body collisional loss rate is determined
by the inelastic scattering amplitude $f_{\beta\alpha}$ for $E_{{\rm s}}=E_{\alpha}$,  i.e., the amplitude of the threshold inelastic scattering. As was shown in Sec. II, when the open channels $\alpha$ and $\beta$ are coupled
to an isolated bound state $|\Phi_{b}\rangle$ with energy $\epsilon_{b}$,
this scattering amplitude can be suppressed to zero, provided that
the condition (\ref{cc}) with $E_{{\rm s}}=E_{\alpha}$ is satisfied.
With straightforward calculation, we find that this condition can
be re-expressed as
\begin{eqnarray}
\epsilon_{b}=\epsilon_{b}^{\ast} & \equiv & E_{\alpha}-\langle\Phi_{b}|WG^{(\alpha\beta)}(E_{\alpha})W|\Phi_{b}\rangle\nonumber \\
 &  & -\frac{2\pi^{2}\langle\Phi_{b}|W|\Psi_{0,\alpha}^{(\alpha\beta+)}\rangle\langle\Phi_{b}|W|\Psi_{\sqrt{E_{\alpha}-E_{\beta}},\beta}^{(\alpha\beta+)}\rangle}{f_{\beta\alpha}^{(\alpha\beta)}(E_{\alpha})}.\nonumber \\
\label{d}
\end{eqnarray}
When Eq. (\ref{d}) is satisfied, the two-body collisional loss is completely
suppressed.

On the other hand, the interaction between two ultracold atoms
in state $|\alpha\rangle_{I}$ can be described by the real part of the
scattering length $a(\epsilon_{b})$, which is defined as
\begin{equation}
a(\epsilon_{b})\equiv-f_{\alpha\alpha}(E_{\alpha},\epsilon_{b}).\label{a}
\end{equation}
Substituting Eq. (\ref{d}) into Eq. (\ref{fji}) and using the optical
theorem, we find that under the conditions of Eq. (\ref{d})
we have
\begin{eqnarray}
{\rm Re}\left[a(\epsilon_{b}^{\ast})\right] & = & a^{({\rm bg})}+f_{\beta\alpha}^{(\alpha\beta)}(E_{\alpha})\frac{\langle\Phi_{b}|W|\Psi_{0,\alpha}^{(\alpha\beta+)}\rangle}{\langle\Phi_{b}|W|\Psi_{\sqrt{E_{\alpha}-E_{\beta}},\beta}^{(\alpha\beta+)}\rangle};\nonumber \\
\label{taa2}\\
{\rm Im}\left[a(\epsilon_{b}^{\ast})\right] & = & 0,
\end{eqnarray}
where $a^{({\rm bg})}=-f_{\alpha\alpha}^{(\alpha\beta)}(E_{\alpha})$
is the scattering length in the system with $W=0$. Eq. (\ref{taa2})
implies that when the inelastic collision is completely suppressed,
the scattering length $a(\epsilon_{b}^{\ast})$ still depends on
details of the two-atom interaction potential $V(r)$ via the factors
$\langle\Phi_{b}|W|\Psi_{0,\alpha}^{(\alpha\beta+)}\rangle$ and $\langle\Phi_{b}|W|\Psi_{\sqrt{E_{\alpha}-E_{\beta}},\beta}^{(\alpha\beta+)}\rangle$. In principle, it is possible for the value of
${\rm Re}\left[a(\epsilon_{b}^{\ast})\right]$ to be much larger than the range $r_{\ast}$ of $V(r)$
(e.g., the van der Waals length), or comparable to $r_{\ast}$, or
much smaller than $r_{\ast}$. When ${\rm Re}\left[a(\epsilon_{b}^{\ast})\right]$ is much larger than the range
$r_{\ast}$, the two atoms in channel $\alpha$ have a large probability
to be close to each other. Nevertheless, because of quantum interference
between the $\alpha\rightarrow\beta$ and $\eta\rightarrow\beta$
transitions, the atoms do not decay to channel $\beta$. In this case,
the interaction between the two atoms in channel $\alpha$
is still strong, while the collisional loss is completely suppressed.

We illustrate our result with the square-well model in Sec. II. D.
Figure 3(a--c) shows the absolute value of the inelastic scattering
amplitude $f_{\beta\alpha}(E_{{\rm s}}=E_{\alpha})$ and the real
part of the scattering length $a(\epsilon_{b}^\ast)$ as functions of the
potential energy $U_{\eta\eta}$ of the closed channel $\eta$, for
three typical cases with the potential energy $U_{\alpha\alpha}$
of channel $\alpha$ taking the values $U_{\alpha\alpha}=-22.5/b^{2}$, $-21.779/b^2$,
and $-21.776/b^{2}$. It is shown that in these three cases, when $|f_{\beta\alpha}|$
is suppressed to zero, the scattering length could be either comparable
or much larger than the range $b$ of the interaction potential.
Figure 3(d) shows the scattering length $a(\epsilon_{b}^{\ast})$
as a function of $U_{\alpha\alpha}$. It is clearly shown that for
systems with different interaction potentials, the value of $a$ ranges from $-\infty$ to $+\infty$.

\begin{figure*}
\includegraphics[width=12cm]{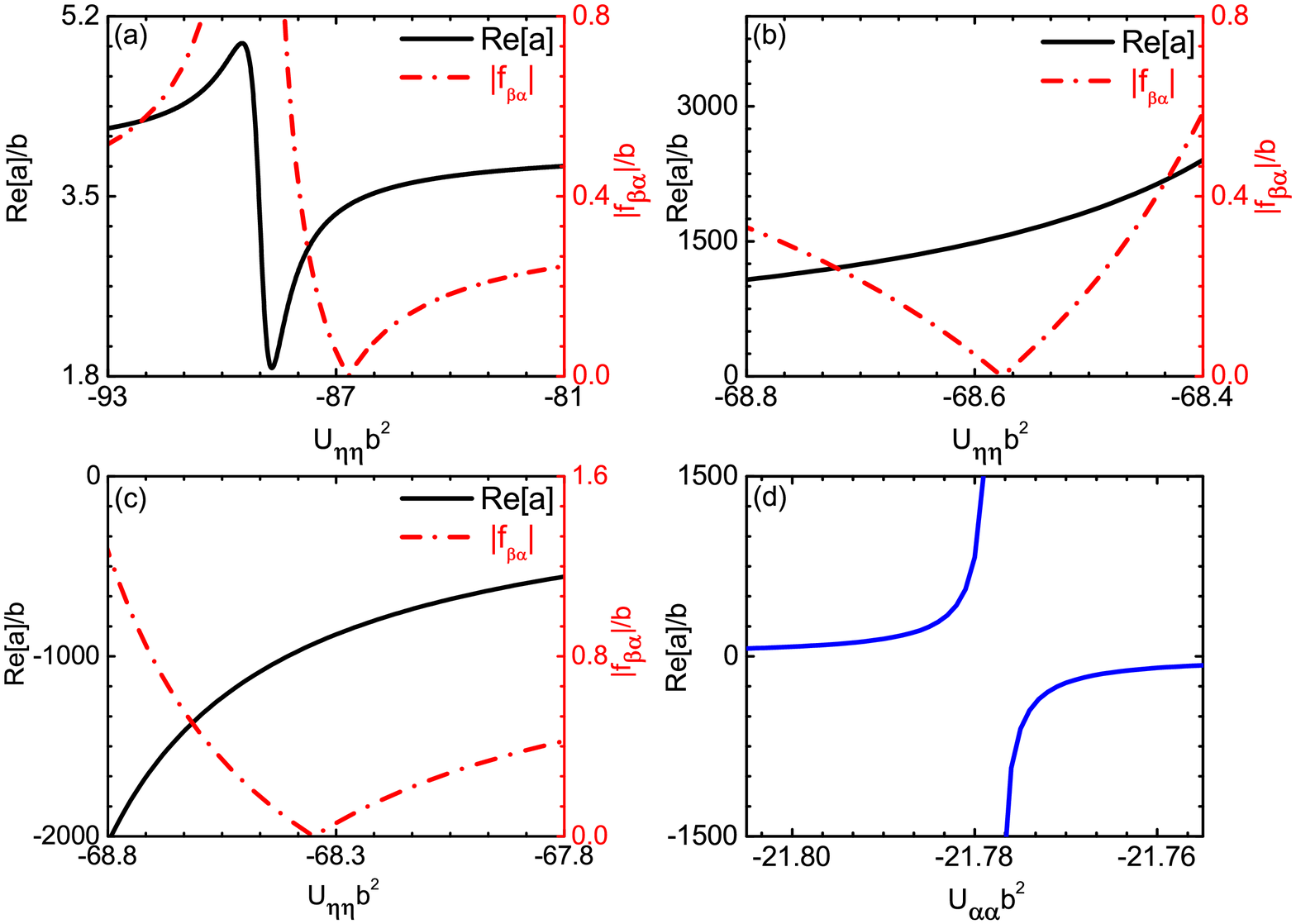}
\caption{(color online) (a-c): The absolute value of the inelastic scattering
amplitude $f_{\beta\alpha}(E_{{\rm s}}=E_{\alpha})$ and the real
part of the scattering length as functions of
the potential energy $U_{\eta\eta}$ of the closed channel $\eta$,
for the square-well model in Sec. II. D. The potential energy
$U_{\alpha\alpha}$ of channel $\alpha$ has the values $U_{\alpha\alpha}=-22.5/b^{2}$
(a), $-21.779/b^{2}$ (b), and $-21.776/b^{2}$ (c). In these three cases, when $|f_{\beta\alpha}|$
is suppressed to zero we have ${\rm Re}[a(\epsilon_{b}^{\ast})]=3.4b$ (a), $1541b$ (b), and $-931b$ (c). (d): The scattering length $a(\epsilon_{b}^{\ast})$
as a function of $U_{\alpha\alpha}$. Our calculation is done with
the parameters $E_{\alpha}=0$, $E_{\beta}=-0.5/b^{2}$, $U_{\beta\beta}=U_{\alpha\beta}=U_{\beta\eta}-3/b^{2}, and U_{\alpha\eta}=2/b^{2}$.}
\end{figure*}

Now we investigate possible experimental realizations of our scheme.
In ultracold gases of alkali atoms, the states $|\beta\rangle_{I}$,
$|\alpha\rangle_{I}$, and $|\eta\rangle_{I}$ can be chosen as the
lowest, second lowest, and higher two-atom hyperfine states with
the same total magnetic quantum number $m_{F}^{(1)}+m_{F}^{(2)}$.
Here $m_{F}^{(1(2))}$ is the magnetic quantum number of atom $1(2)$.
For instance, for ultracold $^{6}$Li atoms, one can choose
\begin{eqnarray}
|\alpha\rangle_{I}\!& = &\!\frac{1}{\sqrt{2}}\left[|\frac{1}{2},-\frac{1}{2}\rangle_{1}|\frac{3}{2},\frac{1}{2}\rangle_{2}-|\frac{3}{2},\frac{1}{2}\rangle_{1}|\frac{1}{2},-\frac{1}{2}\rangle_{2}\right]\!\!;\label{aa}\\
|\beta\rangle_{I} \!& = &\! \frac{1}{\sqrt{2}}\left[|\frac{1}{2},\frac{1}{2}\rangle_{1}|\frac{1}{2},-\frac{1}{2}\rangle_{2}-|\frac{1}{2},-\frac{1}{2}\rangle_{1}|\frac{1}{2},\frac{1}{2}\rangle_{2}\right]\!\!;\label{bb}\\
|\eta\rangle_{I} \!& = &\! \frac{1}{\sqrt{2}}\left[|\frac{3}{2},-\frac{1}{2}\rangle_{1}|\frac{3}{2},\frac{1}{2}\rangle_{2}-|\frac{3}{2},-\frac{1}{2}\rangle_{1}|\frac{3}{2},\frac{1}{2}\rangle_{2}\right]\!\!,\label{cc-1}
\end{eqnarray}
where $|c,d\rangle_{i}$ is the hyperfine state of the $i$-th atom
with $F=c$ and $m_{F}=d$.

Since the total magnetic quantum numbers of three hyperfine channels
$\alpha$, $\beta$, and $\eta$ are the same, these three channels are
coupled to each other via the hyperfine spin-exchange interaction.
Therefore, when we prepare the atoms in channel $|\alpha\rangle_{I}$
(e.g., prepare the ultracold $^{6}$Li atoms in the hyperfine states
$|\frac{1}{2},-\frac{1}{2}\rangle$ and $|\frac{3}{2},\frac{1}{2}\rangle$)
and the threshold energy $E_{\alpha}$ of channel $\alpha$ is
near resonant to a bound state in channel $\eta$, a system shown
in Fig. 1(a) can be realized.

In our system, the threshold energies $E_{\alpha,\beta}$ of channels
$\alpha$, $\beta$, and the energy $\epsilon_{b}$ of the bound state
in channel $\eta$ can be controlled by a static magnetic field via the
Zeeman effect. Therefore, the collisional loss of
atoms in channel $\alpha$ can be suppressed by tuning the magnetic field such that
the condition (\ref{d}) is satisfied. When collisional
loss is suppressed, the elastic scattering length between two
atoms is determined by the details of the inter-atomic interactions, and can be either large or small.

One can also couple the open channel $\alpha$ or $\beta$ and the
bound state $|\Phi_b\rangle$ in a closed hyperfine channel
using a microwave field. In this way it is possible to effectively control the bound-state energy
$\epsilon_{b}$ by changing the frequency of that microwave field \cite{mfr}.
In this case, the total magnetic quantum number of state $|\eta\rangle_{I}$ would differ from that of
 states $|\alpha\rangle_{I}$
and $|\beta\rangle_{I}$  \cite{mfr}.
In addition, in ultracold gases of alkali atoms or alkali-earth (like) atoms,
a laser beam can be used to couple the open scattering channels and a bound state
where one atom is in the electronic ground state and the other atom is in the
electronic excited state \cite{ofr,ofrl}. However,
in this system the spontaneous emission of the excited atom can also
induce atomic losses. As a result, the two-body loss rate can no longer
be suppressed to zero.

\section{Independent control of elastic and inelastic collisions between two
ultracold atoms}

In the preceding sections, we studied the suppression of two-body collisional losses
of ultracold atoms via the Fano effect. We show that when the collisional loss is completely
suppressed, it is still possible for the two-atom scattering length, i.e.,
the threshold elastic scattering amplitude, to be either
large or small. Nevertheless, in that system there is only one control parameter,
i.e., the energy $\epsilon_{b}$ of the isolated bound state. As a result,
when the collisional loss is suppressed by tuning this bound-state energy
to some particular value, the scattering length of these
two atoms would also be entirely fixed, and cannot be altered.

In this section, we study the \emph{independent} control of elastic
and inelastic collisions between two ultracold atoms. To this end,
we first consider the four-channel model shown in Fig. 1(d), where the
IC and OC of the inelastic scattering process
are coupled to two isolated bound states,
rather than a single bound state. We
show that, in this ``ideal'' model,
when the collisional loss of
two atoms in the IC is suppressed by the Fano effect,
the scattering length can still be tuned over a very
broad region by changing the energies of the two bound states. At the end of this section we will discuss a possible
experimental realization of this model.

In the model shown in Fig. 1(d), there are two bound states,
$|\Phi_{b}\rangle$ and $|\Phi_{b^\prime}\rangle$, with energies $\epsilon_{b}$
and $\epsilon_{b'}$, which are located in the closed channels $\eta$ and $\eta^\prime$, respectively.
Each bound state is coupled to the IC $\alpha$ or the OC $\beta$, or both of these two open channels.
It is clear that, in this system,
the two-atom scattering amplitude $f_{jl}$ ($l,j=\alpha,\beta$) from channel $l$ to channel $j$
depends on both of the bound-state energies $\epsilon_{b}$
and $\epsilon_{b'}$, i.e., we have $f_{jl}=f_{jl}[E_{{\rm s}},\epsilon_{b},\epsilon_{b'}]$.
This scattering amplitude can be calculated with the method in Sec.
II. Notice that in the calculation we should replace the channel $\alpha$
in Sec. II with both the channel $\alpha$ and the bound state
$|\Phi_{b'}\rangle$ in our current system.
This straightforward calculation shows that because of the Fano effect,
for any given value of the energy $\epsilon_{b'}$ of the
bound state $|\Phi_{b'}\rangle$, the threshold inelastic collision
can always be completely suppressed if the energy $\epsilon_{b}$
of the bound state $|\Phi_{b}\rangle$ takes a particular $\epsilon_{b'}$-dependent
special value $\chi(\epsilon_{b'})$, i.e., we have
\begin{equation}
f_{\beta\alpha}[E_{\alpha},\epsilon_{b}=\chi(\epsilon_{b'}),\epsilon_{b'}]=0.\label{fba2}
\end{equation}
Furthermore, when $\epsilon_{b}=\chi(\epsilon_{b'})$, the scattering
length $a$ between the two atoms becomes real, and can be expressed
as a function $\bar{a}(\epsilon_{b'})\equiv-f_{\alpha\alpha}[E_{\alpha},\epsilon_{b}=\chi(\epsilon_{b'}),\epsilon_{b'}]$
of the energy $\epsilon_{b'}$. According to the direct calculation
shown in Appendix E, we have
\begin{equation}
\bar{a}(\epsilon_{b'})=a^{(\alpha\beta\eta')}+\frac{A'}{E_{\alpha}-\epsilon_{b'}-B'},\label{chi-2}
\end{equation}
where $a^{(\alpha\beta\eta')}$ is the scattering length in the system
with $V_{\beta\eta}=V_{\alpha\eta}=0$. The expressions of the parameters
$A'$ and $B'$ are given in Appendix E. In this appendix we also
prove that $B'$ is a $\epsilon_{b'}$-independent real parameter.
Because of this and considering Eq. (\ref{chi-2}), $\bar{a}(\epsilon_{b'})$ can be controlled in a very broad region by tuning the bound-state energy
$\epsilon_{b'}$ in the region around $E_{\alpha}-B'$.

Here we illustrate this control effect using calculations with a square-well potential. In our
model, the total Hamiltonian is ${\bf p}^{2}+\sum_{j=\alpha,\beta,\eta,\eta'}E_{j}|j\rangle_{I}\langle j|+\sum_{l,j=\alpha,\beta,\eta,\eta'}V_{lj}(r)|l\rangle_{I}\langle j|$,
where $V_{lj}(r)=U_{lj}$ for $r<b$, $V_{lj}(r)=0$ for $r>b$, $E_{\eta}=E_{\eta'}=\infty$,
$E_{\alpha}=0$, and $E_{\beta}<0$. In Fig. 4 we illustrate the scattering
length $\bar{a}(\epsilon_{b'})$ as a function of the energy $\epsilon_{b'}$
of the lowest bound state in channel $\eta'$. It is clearly shown
that this scattering length can be resonantly controlled by $\epsilon_{b'}$
or the potential energy $U_{\eta\prime\eta\prime}$ of channel $\eta\prime$.

\begin{figure}
\includegraphics[width=8cm]{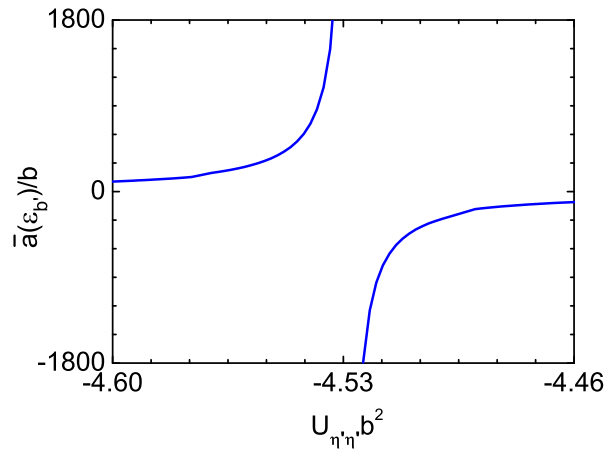}
\caption{The elastic scattering length ${\bar a}(\epsilon_{b'})\equiv-f_{\alpha\alpha}[E_{\alpha},\epsilon_{b}=\chi(\epsilon_{b'}),\epsilon_{b'}]$
for the square-well model in Sec. IV. Here $\epsilon_{b'}$ is the
lowest bound state in channel $\eta\prime$. The calculation is executed
with the parameters $E_{\alpha}=0$, $E_{\beta}=-0.5/b^{2}$,
$U_{\alpha\alpha}=-1.5/b^{2}, U_{\beta\beta}=-3/b^{2}$,
$U_{\alpha\beta}=U_{\beta\eta}=U_{\alpha\eta'}=3/b^{2}, and U_{\alpha\eta}=U_{\eta\eta'}=U_{\beta\eta'}=0$.}
\end{figure}

Here, we propose one possible experimental realization
of the model discussed in this section. In an ultracold gas of alkali
atoms, the states $|l\rangle_{I}$ ($l=\alpha,\beta,\eta,\eta'$)
can be chosen as a two-atom hyperfine state $|l\rangle_{I}$, which
satisfies $(F_{z}^{(1)}+F_{z}^{(2)})|l\rangle_{I}=M_{l}|l\rangle_{I}$.
Under the condition, $M_{\alpha}=M_{\beta}=M_{\eta}$, the channels
$\alpha$, $\beta$, and $\eta$ are coupled via hyperfine interactions.
Aided by the Zeeman effect, the energy $\epsilon_{b}$ of the
bound state $|\Phi_{b}\rangle$ in channel $\eta$ can be controlled
by a static magnetic field. In addition, with a microwave field one
can further couple the open channels $\alpha$ and $\beta$ with the
bound state $|\Phi_{b'}\rangle$ in channel $\eta'$, and effectively
control the energy $\epsilon_{b'}$ by altering the frequency of that microwave
field.

\section{Summary}

In this paper we generalize the Fano effect to systems with arbitrary
IC--OC coupling strengths. We prove that in systems with one IC and one OC,
when
the inter-atomic interaction potential is real,
the $s$-wave inelastic scattering amplitude can always be suppressed to zero by the coupling
between these open channels and an isolated bound state. Using our result, we further
show that when the two-body collisional loss of an ultracold gas is suppressed
via the Fano effect,
 it is possible for the two-atom elastic scattering length
to be either much larger, comparable to or smaller than the van der Waals length.
We also show that when the open channels are coupled to two bound
states, the elastic scattering
length of the atoms in the higher open channel can be resonantly controlled, while
the inelastic scattering is completely suppressed. Our results show that the Fano effect
may be a very powerful technique for the suppression of collisional losses in ultracold gases. Furthermore,
the generalized Fano effect we derived in Sec. II may also be useful
for the study of the inelastic scattering processes in other systems.

It is pointed out that in this paper we consider systems with
spherically symmetrical interaction potentials. Nevertheless, the Fano effect can also
be used to suppress the collisional losses induced by anisotropic interactions, e.g.,
dipolar losses caused by dipole-dipole interactions \cite{Rb85loss2,Rb85loss3}.
In these cases, although the collisional losses cannot be suppressed to zero, they can also be significantly decreased (e.g., decreased by more than one order of magnitude \cite{Rb85loss2,Rb85loss3})
when one or several open channels are coupled to an isolated bound state.

\begin{acknowledgments}
This work has been supported by the National Natural Science Foundation
of China under Grant Nos. 11222430 and 11434011, and by NKBRSF of China
under Grant No. 2012CB922104. Peng Zhang also thanks Hui Zhai and T. L. Ho for
helpful discussions.
\end{acknowledgments}
\appendix
\addcontentsline{toc}{section}{Appendices}\markboth{APPENDICES}{}
\begin{subappendices}

\section{Proof of Eq. (\ref{psis}) }

In this appendix we prove Eq. (\ref{psis}). According to the formal
scattering theory, the scattering state $|\Psi_{k_{l},l}^{(+)}\rangle$
satisfies the equation \cite{fb}
\begin{equation}
|\Psi_{k_{l},l}^{(+)}\rangle=\lim_{\lambda\rightarrow0^{+}}\frac{i\lambda}{E_{{\rm s}}+i\lambda-H}|\Psi_{k_{l},l}^{(0)}\rangle,\label{lp1}
\end{equation}
where $E_{{\rm s}}$ and $H$ are defined in Eq. (\ref{se-1}) and
Eq. (\ref{hab}), respectively, and the state $|\Psi_{k_{l},l}^{(0)}\rangle$
is defined in Sec. II. A. Similarly, the state $|\Psi_{k_{l},l}^{(\alpha\beta+)}\rangle$,
which is the $s$-wave component of the out-going scattering state
for the case with $W=0$, satisfies the equation
\begin{equation}
|\Psi_{k_{l},l}^{(\alpha\beta+)}\rangle=\lim_{\lambda\rightarrow0^{+}}\frac{i\lambda}{E_{{\rm s}}+i\lambda-(H-W)}|\Psi_{k_{l},l}^{(0)}\rangle.\label{lp1-1}
\end{equation}
Substituting the relation
\begin{eqnarray}
 &  & \frac{1}{E_{{\rm s}}+i\lambda-H}=\frac{1}{E_{{\rm s}}+i\lambda-(H-W)}\nonumber \\
 &  & +\frac{1}{E_{{\rm s}}+i\lambda-(H-W)}W\frac{1}{E_{{\rm s}}+i\lambda-H}\label{g1}
\end{eqnarray}
into Eq. (\ref{lp1}), and using Eq. (\ref{lp1-1}) and Eq. (\ref{gab}),
we can obtain Eq. (\ref{psis}).

\section{Proof of Eq. (\ref{fji})}

In this appendix we prove Eq. (\ref{fji}). To this end, we substitute
Eq. (\ref{gg2}) into Eq. (\ref{psis}). Then we find that the solution
of Eq. (\ref{psis}) can be expressed
\begin{equation}
|\Psi_{k_{l},l}^{(+)}\rangle=|\Gamma_{k_{l},l}\rangle+\kappa|\Phi_{b}\rangle,\label{psirw}
\end{equation}
where the state $|\Gamma_{k_{l},l}\rangle$ is in the subspace spanned
by $|\alpha\rangle_{I}$ and $|\beta\rangle_{I}$, and $\kappa$ is
a c-number. Furthermore, using Eq. (\ref{gg2}) we can rewrite Eq.
(\ref{psis}) as the equations of $|\Gamma_{k_{l},l}\rangle$ and
$\kappa$:
\begin{eqnarray}
|\Gamma_{k_{l},l}\rangle & = & |\Psi_{k_{l},l}^{(\alpha\beta+)}\rangle+\kappa G^{(\alpha\beta)}(E_{{\rm s}})W|\Phi_{b}\rangle,\label{kappa}\\
\kappa & = & \frac{\langle\Phi_{b}|W|\Gamma_{k_{l},l}\rangle}{E_{{\rm s}}-\epsilon_{b}}.\label{k2-1}
\end{eqnarray}
Substituting Eq. (\ref{kappa}) into Eq. (\ref{k2-1}), we obtain
the equation
\begin{equation}
\kappa=\frac{\langle\Phi_{b}|W|\Psi_{k_{l},l}^{(\alpha\beta+)}\rangle}{E_{{\rm s}}-\epsilon_{b}}+\kappa\frac{\langle\Phi_{b}|WG^{(\alpha\beta)}(E_{{\rm s}})W|\Phi_{b}\rangle}{E_{{\rm s}}-\epsilon_{b}},\label{ka2}
\end{equation}
which gives
\begin{equation}
\kappa=\frac{\langle\Phi_{b}|W|\Psi_{k_{l},l}^{(\alpha\beta+)}\rangle}{E_{{\rm s}}-\epsilon_{b}-\langle\Phi_{b}|WG^{(\alpha\beta)}(E_{{\rm s}})W|\Phi_{b}\rangle}.\label{ka3}
\end{equation}
Substituting this result into Eq. (\ref{kappa}), we can further derive
the state $|\Gamma_{k_{l},l}\rangle$.

Using these results, we can calculate the scattering amplitude $f_{jl}(E_{{\rm s}},\epsilon_{b})$.
Substituting Eq. (\ref{psirw}) into Eq. (\ref{f}), we obtain
\begin{eqnarray}
f_{jl}(E_{{\rm s}},\epsilon_{b}) & = & -2\pi^{2}\langle\Psi_{k_{j},j}^{(0)}|(V-W)|\Gamma_{k_{l},l}\rangle\nonumber \\
 &  & -2\pi^{2}\kappa\langle\Psi_{k_{j},j}^{(0)}|W|\Phi_{b}\rangle.\label{fa1}
\end{eqnarray}
Substituting Eqs. (\ref{ka3}, \ref{kappa}) into Eq. (\ref{fa1}),
and using the relation
\begin{equation}
f_{jl}^{(\alpha\beta)}(E_{{\rm s}})=-2\pi^{2}\langle\Psi_{k_{j},j}^{(0)}|(V-W)|\Psi_{k_{l},l}^{(\alpha\beta+)}\rangle\label{fa2-1}
\end{equation}
satisfied by the scattering amplitude $f_{jl}^{(\alpha\beta)}(E_{{\rm s}})$
for the case with $W=0$, we obtain
\begin{eqnarray}
f_{jl}(E_{{\rm s}},\epsilon_{b}) & = & f_{jl}^{(\alpha\beta)}(E_{{\rm s}})\nonumber \\
 &  & -2\pi^{2}\frac{\langle\Psi_{k_{j},j}^{(\alpha\beta-)}|W|\Phi_{b}\rangle\langle\Phi_{b}|W|\Psi_{k_{l},l}^{(\alpha\beta+)}\rangle}{E_{{\rm s}}-\epsilon_{b}-\langle\Phi_{b}|WG^{(\alpha\beta)}(E_{{\rm s}})W|\Phi_{b}\rangle}.\nonumber \\
\label{fjl3}
\end{eqnarray}
Here $|\Psi_{k_{j},j}^{(\alpha\beta-)}\rangle$ is the $s$-wave
component of the incoming scattering state for the case with $W=0$,
with respect to incident channel $j$ and incident momentum ${\bf k}_{j}$.
It satisfies the Lippman--Schwinger equation
\begin{equation}
|\Psi_{k_{j},j}^{(\alpha\beta-)}\rangle=|\Psi_{k_{j},j}^{(0)}\rangle+G^{(\alpha\beta)}(E_{{\rm s}})^{\dagger}(V-W)|\Psi_{k_{j},j}^{(0)}\rangle\label{lpse}
\end{equation}
and the relation
\begin{equation}
_{I}\langle\l|_{R}\langle{\bf r}|\Psi_{k_{j},j}^{(\alpha\beta-)}\rangle=\ _{I}\langle\l|_{R}\langle{\bf r}|\Psi_{k_{j},j}^{(\alpha\beta+)}\rangle^{\ast}\label{rl}
\end{equation}
for $l=\alpha,\ \beta$. Here $|{\bf r}\rangle_{R}$ is the eigen-state of the relative position
of the two atoms. Because of the relation (\ref{rl}), we have
\begin{equation}
\langle\Psi_{k_{j},j}^{(\alpha\beta-)}|W|\Phi_{b}\rangle=\langle\Phi_{b}|W|\Psi_{k_{j},j}^{(\alpha\beta+)}\rangle.\label{rl2}
\end{equation}
Substituting Eq. (\ref{rl2}) into Eq. (\ref{fjl3}), we can obtain
Eq. (\ref{fji}).

\section{$S$-matrix in the $s$-wave subspace}

In this appendix we prove some properties of the $S$-matrix related
to the $s$-wave scattering in our system, which is introduced in
Sec. II. B.

We first study the relation between this $S$-matrix and the $S$-operator
in our system. To this end, we introduce a state $|\Phi_{k,l}^{(0)}\rangle$
($l=\alpha,\beta$), which is defined as $|\Phi_{k,l}^{(0)}\rangle=\sqrt{2\pi k}|\Psi_{k,l}^{(0)}\rangle$.
Here $|\Psi_{k,l}^{(0)}\rangle$ is defined in Sec. II. A. It is easy
to prove that
\begin{equation}
_{R}\langle{\bf r}|\Phi_{k,l}^{(0)}\rangle=\frac{\sin(kr)}{2\pi\sqrt{k}r}|l\rangle_{I}.\label{ac1}
\end{equation}
This relation yields \cite{st}
\begin{equation}
\langle\Phi_{k',l'}^{(0)}|\Phi_{k,l}^{(0)}\rangle=\delta_{l,l'}\delta(E_{k,l}-E_{k',l'})\label{zj}
\end{equation}
and
\begin{equation}
\sum_{l}\int dE|\Phi_{\sqrt{E-E_{l}},l}^{(0)}\rangle\langle\Phi_{\sqrt{E-E_{l}},l}^{(0)}|=1,\label{wan}
\end{equation}
where the energy $E_{k,l}$ ($l=\alpha,\beta$) is defined as $E_{k,l}=k^{2}+E_{l}.$

Now let us consider the factor $\langle\Phi_{k',l'}^{(0)}|\hat{S}|\Phi_{k,l}^{(0)}\rangle$
($l,l'=\alpha,\beta$), where $\hat{S}$ is the $S$-operator of our
system. It is defined as $\hat{S}=\Omega_{-}^{\dagger}\Omega_{+}$,
where $\Omega_{\pm}$ are the M$\mathrm{\phi}$ller operators \cite{st}.
According to the formal scattering theory \cite{st}, we have
\begin{equation}
\langle\Phi_{k',l'}^{(0)}|\hat{S}|\Phi_{k,l}^{(0)}\rangle=\langle\Phi_{k',l'}^{(-)}|\Phi_{k,l}^{(+)}\rangle,\label{ss}
\end{equation}
with
\begin{equation}
|\Phi_{k,l}^{(\pm)}\rangle=\sqrt{2\pi k}|\Psi_{k,l}^{(\pm)}\rangle\label{pp}
\end{equation}
($l=\alpha,\beta$). Here $|\Psi_{k,l}^{(+/-)}\rangle$ is the $s$-wave
component of the incoming/out-going scattering state with scattering
energy $E_{k,l}$ and incident channel $l$, as defined in Sec. II. A. They satisfy the Lippman--Schwinger
equation
\begin{equation}
|\Psi_{k,l}^{(\pm)}\rangle=|\Psi_{k,l}^{(0)}\rangle+\lim_{\lambda\rightarrow0^{+}}\frac{1}{E_{k,l}\pm i\lambda-H}V|\Psi_{k,l}^{(0)}\rangle,\label{lse3}
\end{equation}
the Schr$\ddot{{\rm o}}$dinger equation $H|\Psi_{k,l}^{(\pm)}\rangle=E_{k,l}|\Psi_{k,l}^{(\pm)}\rangle$, and
the normalization condition $\langle\Psi_{k',l'}^{(+)}|\Psi_{k,l}^{(+)}\rangle=\langle\Psi_{k',l'}^{(-)}|\Psi_{k,l}^{(-)}\rangle=\langle\Psi_{k',l'}^{(0)}|\Psi_{k,l}^{(0)}\rangle.$
These facts yield
\begin{eqnarray}
 &  & |\Phi_{k',l'}^{(-)}\rangle=|\Phi_{k',l'}^{(+)}\rangle+\label{sse}\\
 &  & \lim_{\lambda\rightarrow0^{+}}\left(\frac{1}{E_{k',l'}-i\lambda-H}-\frac{1}{E_{k',l'}+i\lambda-H}\right)V|\Phi_{k',l'}^{(0)}\rangle,\nonumber \\
 &  & H|\Phi_{k,l}^{(\pm)}\rangle=E_{k,l}|\Phi_{k,l}^{(\pm)}\rangle,\label{se-1}\\
 &  & \langle\Phi_{k',l'}^{(+)}|\Phi_{k,l}^{(+)}\rangle=\langle\Phi_{k',l'}^{(-)}|\Phi_{k,l}^{(-)}\rangle=\delta(E_{k',l'}-E_{k,l})\delta_{l,l^{\prime}}.\label{nc}
\end{eqnarray}
Substituting Eq. (\ref{sse}) into Eq. (\ref{ss}), and using Eq.
(\ref{nc}) and (\ref{se-1}), we obtain
\begin{eqnarray}
\langle\Phi_{k',l'}^{(0)}|\hat{S}|\Phi_{k,l}^{(0)}\rangle & = & \delta(E_{k',l'}-E_{k,l})\delta_{l,l^\prime}+\nonumber \\
 &  & \left(\frac{1}{x+i0^{+}}-\frac{1}{x-i0^{+}}\right)\langle\Phi_{k',l'}^{(0)}|V|\Phi_{k,l}^{(+)}\rangle,\nonumber \\
\label{ee}
\end{eqnarray}
where $x=E_{k',l'}-E_{k,l}$. With the help of the relation
\begin{equation}
\frac{1}{x+i0^{+}}-\frac{1}{x+i0^{-}}=-2\pi i\delta(x),\label{re}
\end{equation}
and Eqs. (\ref{f}) and (\ref{pp}), we can further rewrite Eq. (\ref{ee})
as
\begin{equation}
\langle\Phi_{k',l'}^{(0)}|\hat{S}|\Phi_{k,l}^{(0)}\rangle=\delta(E_{k',l'}-E_{k,l})S_{l',l}(E_{k,l}),\label{ss2}
\end{equation}
where $S_{l',l}(E_{k,l})$ is defined in Eq. (\ref{fs}). This is
the relation between the $S$-matrix and the $S$-operator $\hat{S}$
in our system.

Now we prove the $S$-matrix is a unitary matrix. Since the $S$-operator
$\hat{S}$ is a unitary operator \cite{st}, it satisfies
\begin{equation}
\hat{S}^{\dagger}\hat{S}=1.\label{su}
\end{equation}
Using this result and Eqs. (\ref{zj}) and (\ref{wan}), we obtain
\begin{eqnarray}
 &  & \sum_{l''}\int dE''\langle\Phi_{\sqrt{E''-E_{l''}},l''}^{(0)}|\hat{S}|\Phi_{k',l'}^{(0)}\rangle^{\ast}\langle\Phi_{\sqrt{E''-E_{l''}},l''}^{(0)}|\hat{S}|\Phi_{k,l}^{(0)}\rangle\nonumber \\
 &  & =\delta_{l,l'}\delta(E_{k',l'}-E_{k,l}).\label{un}
\end{eqnarray}
Substituting Eq. (\ref{ss2}) into Eq. (\ref{un}), we find that the
$2\times2$ matrix with element $S_{l',l}(E)$, i.e., the $S$-matrix
we introduced in Eq. (\ref{s}), is a unitary matrix.

Now we consider the $S$-matrix in the system with real interaction potential. In such a system, the $S$-operator $\hat{S}$ satisfies
\cite{st}
\begin{eqnarray}
\langle\Phi^\prime|\hat{S}|\Phi\rangle=\langle\widetilde{\Phi}|\hat{S}|\widetilde{\Phi}^\prime\rangle,
\end{eqnarray}
where the state $|\widetilde{\Phi}\rangle$ is defined
as $|\widetilde{\Phi}\rangle={\cal T}|\Phi\rangle$, with ${\cal T}$
the time-reversal operator for the spatial motion. The state $|\widetilde{\Phi}\rangle$
satisfies the relation \cite{st}
\begin{equation}
_{R}\langle{\bf r}|\widetilde{\Phi}\rangle={}_{R}\langle{\bf r}|\Phi\rangle^{\ast}.\label{tr}
\end{equation}
From Eqs. (\ref{ac1}) and (\ref{tr}), we know that $|\widetilde{\Phi}_{k,l}^{(0)}\rangle=|\Phi_{k,l}^{(0)}\rangle$
for $l=\alpha,\beta$. Therefore, we have
\begin{equation}
\langle\Phi_{k',l'}^{(0)}|\hat{S}|\Phi_{k,l}^{(0)}\rangle=\langle\Phi_{k,l}^{(0)}|\hat{S}|\Phi_{k',l'}^{(0)}\rangle.\label{tr2}
\end{equation}
This result and the relation (\ref{ss2}) indicates that the $S$-matrix
defined in Eq. (\ref{s}) is a symmetric matrix for the system with real potentials.

\section{The ratio $D_{1}(E_{{\rm s}})/C_{1}(E_{{\rm s}})$}

In this appendix we prove that the ratio $D_{1}(E_{{\rm s}})/C_{1}(E_{{\rm s}})$
appearing in Sec. II. C is real. To this end, we will first prove that all the ratios
 $C_{1}(E_{{\rm s}})/C_{2}(E_{{\rm s}})$, $D_{1}(E_{{\rm s}})/D_{2}(E_{{\rm s}})$,
 and $D_{2}(E_{{\rm s}})/C_{2}(E_{{\rm s}})$ are real.

According to Eqs. (\ref{c1}) and (\ref{c2}),
the ratio $C_{1}(E_{{\rm s}})/C_{2}(E_{{\rm s}})$ is just the non-diagonal
element of the $K$-matrix for the case with $W=0$. As shown in Sec.
II. C, in our system this matrix
element is real. Thus, $C_{1}(E_{{\rm s}})/C_{2}(E_{{\rm s}})$ is
real.

Moreover, according to Eq. (\ref{kab}), we have $D_{1}(E_{{\rm s}})/D_{2}(E_{{\rm s}})=K_{\beta\alpha}(E_{{\rm s}},0)$.
Since $K_{\beta\alpha}(E_{{\rm s}},\epsilon_{b})$ is real for any
$\epsilon_{b}$, the ratio $D_{1}(E_{{\rm s}})/D_{2}(E_{{\rm s}})$
is also real.

Now we prove that $D_{2}(E_{{\rm s}})/C_{2}(E_{{\rm s}})$ is also
real. We can prove this result by contradiction. To this end, we re-express
Eq. (\ref{kab}) as
\begin{eqnarray}
\left(\frac{C_{2}(E_{{\rm s}})}{C_{1}(E_{{\rm s}})}\right)K_{\beta\alpha}(E_{{\rm s}},\epsilon_{b}) & = & \frac{\epsilon_{b}+\left[\frac{D_{1}(E_{{\rm s}})}{C_{1}(E_{{\rm s}})}\right]}{\epsilon_{b}+\left[\frac{D_{2}(E_{{\rm s}})}{C_{2}(E_{{\rm s}})}\right]}.\nonumber \\
\label{kab2}
\end{eqnarray}
Since both $K_{\beta\alpha}(E_{{\rm s}},\epsilon_{b})$ and $C_{2}(E_{{\rm s}})/C_{1}(E_{{\rm s}})$
are real, the right-hand side of Eq. (\ref{kab2}) is real for any
$\epsilon_{b}$. Thus, if $D_{2}(E_{{\rm s}})/C_{2}(E_{{\rm s}})$
is not real, we must have $\frac{D_{1}(E_{{\rm s}})}{C_{1}(E_{{\rm s}})}=\frac{D_{2}(E_{{\rm s}})}{C_{2}(E_{{\rm s}})}$.
Using Eq. (\ref{kab}), we find that this result yields that $K_{\beta\alpha}(E_{{\rm s}},\epsilon_{b})=C_{1}(E_{{\rm s}})/C_{2}(E_{{\rm s}})$,
i.e., $K_{\beta\alpha}(E_{{\rm s}},\epsilon_{b})$ is \emph{independent}
of $\epsilon_{b}$. Furthermore, with Eqs. (\ref{ks}, \ref{fji},
\ref{fs}) we can express the diagonal elements of the $K$-matrix
as
\begin{eqnarray}
K_{\alpha\alpha}(E_{{\rm s}},\epsilon_{b}) & = & \frac{C_{1}^{\prime}(E_{{\rm s}})\epsilon_{b}+D_{1}^{\prime}(E_{{\rm s}})}{C_{2}(E_{{\rm s}})\epsilon_{b}+D_{2}(E_{{\rm s}})};\label{kaa}\\
K_{\beta\beta}(E_{{\rm s}},\epsilon_{b}) & = & \frac{C_{1}^{\prime\prime}(E_{{\rm s}})\epsilon_{b}+D_{1}^{\prime\prime}(E_{{\rm s}})}{C_{2}(E_{{\rm s}})\epsilon_{b}+D_{2}(E_{{\rm s}})},\label{kbb}
\end{eqnarray}
where the factors $C_{1}^{\prime}(E_{{\rm s}})$, $D_{1}^{\prime}(E_{{\rm s}})$,
$C_{1}^{\prime\prime}(E_{{\rm s}})$, and $D_{1}^{\prime\prime}(E_{{\rm s}})$
are given by\begin{widetext}
\begin{eqnarray}
C_{1}^{\prime}(E_{{\rm s}}) & = & -i\left[1-s_{\alpha\alpha}(E_{{\rm s}})s_{\beta\beta}(E_{{\rm s}})+s_{\alpha\beta}(E_{{\rm s}})s_{\beta\alpha}(E_{{\rm s}})-s_{\alpha\alpha}(E_{{\rm s}})+s_{\beta\beta}(E_{{\rm s}})\right];\label{c1p}\\
D_{1}^{\prime}(E_{{\rm s}}) & = & -C_{1}^{\prime}(E_{{\rm s}})B(E_{{\rm s}})-i\left\{ s_{\alpha\alpha}(E_{{\rm s}}){\cal A}_{\beta\beta}(E_{{\rm s}})+s_{\beta\beta}(E_{{\rm s}}){\cal A}_{\alpha\alpha}(E_{{\rm s}})-{\cal A}_{\alpha\beta}(E_{{\rm s}})\left[s_{\beta\alpha}(E_{{\rm s}})+s_{\alpha\beta}(E_{{\rm s}})\right]\right\} \nonumber \\
 &  & -i{\cal A}_{\alpha\alpha}(E_{{\rm s}})+i{\cal A}_{\beta\beta}(E_{{\rm s}});\label{d1p}\\
C_{1}^{\prime\prime}(E_{{\rm s}}) & = & -i\left[1-s_{\alpha\alpha}(E_{{\rm s}})s_{\beta\beta}(E_{{\rm s}})+s_{\alpha\beta}(E_{{\rm s}})s_{\beta\alpha}(E_{{\rm s}})+s_{\alpha\alpha}(E_{{\rm s}})+s_{\beta\beta}(E_{{\rm s}})\right];\label{c1pp}\\
D_{1}^{\prime\prime}(E_{{\rm s}}) & = & -C_{1}^{\prime\prime}(E_{{\rm s}})B(E_{{\rm s}})-i\left\{ s_{\alpha\alpha}(E_{{\rm s}}){\cal A}_{\beta\beta}(E_{{\rm s}})+s_{\beta\beta}(E_{{\rm s}}){\cal A}_{\alpha\alpha}(E_{{\rm s}})-{\cal A}_{\alpha\beta}(E_{{\rm s}})\left[s_{\beta\alpha}(E_{{\rm s}})+s_{\alpha\beta}(E_{{\rm s}})\right]\right\} \nonumber \\
 &  & i{\cal A}_{\alpha\alpha}(E_{{\rm s}})-i{\cal A}_{\beta\beta}(E_{{\rm s}}).\label{d1pp}
\end{eqnarray}
\end{widetext} Since the $K$-matrix is a Hermitian matrix, $K_{\alpha\alpha}(E_{{\rm s}},\epsilon_{b})$
and $K_{\beta\beta}(E_{{\rm s}},\epsilon_{b})$ are real for any $\epsilon_{b}$.
Therefore, with a similar method to that used above, we find that if $D_{2}(E_{{\rm s}})/C_{2}(E_{{\rm s}})$
is not real, both $K_{\alpha\alpha}(E_{{\rm s}},\epsilon_{b})$ and
$K_{\beta\beta}(E_{{\rm s}},\epsilon_{b})$ are\emph{ }independent
of $\epsilon_{b}$. Therefore, if $D_{2}(E_{{\rm s}})/C_{2}(E_{{\rm s}})$
is not real, all the $K$-matrix elements are independent of $\epsilon_{b}$.
According to Eqs. (\ref{k}) and (\ref{fs}), this result indicates that all
scattering amplitudes $f_{jl}(E_{{\rm s}},\epsilon_{b})$ ($j,l=\alpha,\beta$)
are independent of $\epsilon_{b}$. However, according to Eq. (\ref{fji}),
$f_{jl}(E_{{\rm s}},\epsilon_{b})$ takes different values for different
$\epsilon_{b}$. Therefore, in our system the ratio $D_{2}(E_{{\rm s}})/C_{2}(E_{{\rm s}})$
is real.

So far we have shown that the ratios $C_{1}(E_{{\rm s}})/C_{2}(E_{{\rm s}})$,
$D_{1}(E_{{\rm s}})/D_{2}(E_{{\rm s}})$, and $D_{2}(E_{{\rm s}})/C_{2}(E_{{\rm s}})$
are all real. It follows that the ratio $D_{1}(E_{{\rm s}})/C_{1}(E_{{\rm s}})$
is also real.

\section{Eq. (\ref{chi-2}) and The Parameter $B'$}

In this appendix we will prove Eq. (\ref{chi-2}) in Sec. IV, and
prove that the parameter $B'$ is real in this equation.

\subsection{Proof of Eq. (\ref{chi-2})}

The Hamiltonian for the system in this section is
\begin{eqnarray}
H_{T} & = & H+\left[E_{\eta'}+V_{\eta'\eta'}(r)\right]|\eta'\rangle_{I}\langle\eta'|\nonumber \\
 &  & +\sum_{l=\alpha,\beta}V_{l\eta'}(r)|l\rangle_{I}\langle\eta'|+h.c.\label{ht}
\end{eqnarray}
Here the Hamiltonian $H$ is defined in Eq. (\ref{hab}). It describes
the relative kinetic energy and the interaction potential of two atoms
in channels $\alpha,$ $\beta$, and $\eta$. In Eq. (\ref{ht}), $E_{\eta'}$
and $V_{\eta'\eta'}(r)$ are the threshold energy and interaction
potential of channel $\eta'$, respectively, and $V_{\alpha\eta'}(r)$
is the inter-channel coupling between channel $\alpha$ and $\eta'$.
Here we also assume $V_{\eta'\eta'}(r)$ and $V_{\alpha\eta'}(r)$
are real functions of $r$ and tend to zero in the limit $r\rightarrow0$.

As shown in our main text, for this system we can obtain the scattering
amplitude $f_{jl}[E_{{\rm s}},\epsilon_{b},\epsilon_{b'}]$ with the
method in Sec. II. With this method we find that when the collisional
decay from channel $\alpha$ to channel $\beta$ is completely suppressed,
i.e., under the condition $\epsilon_{b}=\chi(\epsilon_{b'})$, the
scattering length between two ultracold atoms in channel $\alpha$
is given by
\begin{eqnarray}
{\bar a}(\epsilon_{b'}) & \equiv & -f_{\alpha\alpha}[E_{\alpha},\epsilon_{b}=\chi(\epsilon_{b'}),\epsilon_{b'}]\nonumber \\
 & = & a^{(\alpha\beta\eta')}+f_{\beta\alpha}^{(\alpha\beta\eta')}(E_{\alpha})\frac{\langle\Phi_{b}|W|\Psi_{0,\alpha}^{(\alpha\beta\eta'+)}\rangle}{\langle\Phi_{b}|W|\Psi_{\sqrt{E_{\alpha}-E_{\beta}},\beta}^{(\alpha\beta\eta'+)}\rangle},\nonumber \\
\label{a3}
\end{eqnarray}
where the operator $W$ is defined in Eq. (\ref{w-1-1}), and $|\Psi_{k,l}^{(\alpha\beta\eta'+)}\rangle$
($l=\alpha,\beta$) is the $s$-wave component of the out-going scattering
state in the system with $W=0$, with incident momentum $k$ and incident
channel $l$. In Eq. (\ref{a3}) $f_{jl}^{(\alpha\beta\eta')}(E_{{\rm s}})$
($j,l=\alpha,\beta$) is the scattering amplitude for the system with
$W=0$, with incident channel $l$, out-going channel $j$, and scattering
energy $E_{{\rm s}}$, and $a^{(\alpha\beta\eta')}$ is defined as
$a^{(\alpha\beta\eta')}\equiv-f_{\alpha\alpha}^{(\alpha\beta\eta')}(E_{\alpha})$.

Now we calculate the state $|\Psi_{\sqrt{E_{\alpha}-E_{\beta}},\beta}^{(\alpha\beta\eta'+)}\rangle$.
With the method in Appendix A, we can easily prove that the state
$|\Psi_{\sqrt{E_{\alpha}-E_{\beta}},\beta}^{(\alpha\beta\eta'+)}\rangle$
satisfies the equation
\begin{eqnarray}
|\Psi_{\sqrt{E_{\alpha}-E_{\beta}},\beta}^{(\alpha\beta\eta'+)}\rangle & = & |\Psi_{\sqrt{E_{\alpha}-E_{\beta}},\beta}^{(\alpha\beta+)}\rangle\nonumber \\
 &  & +g^{(\alpha\beta)}(E_{\alpha})W'|\Psi_{\sqrt{E_{\alpha}-E_{\beta}},\beta}^{(\alpha\beta\eta'+)}\rangle,\label{lse3-1}
\end{eqnarray}
where $W'=\sum_{l=\alpha,\beta}V_{l\eta'}(r)|l\rangle_{I}\langle\eta'|+h.c.$,
and the operator $g^{(\alpha\beta)}(E)$ is defined as $g^{(\alpha\beta)}(E)=1/[E+i0^{+}-(H_{T}-W-W')]$.
Here, the state $|\Psi_{k,l}^{(\alpha\beta+/-)}\rangle$ ($l=\alpha,\beta$)
is the $s$-wave component of the out-going/incoming scattering state
in the system with $W=W'=0$, with incident momentum $k$ and incident
channel $l$. Similar to that in Sec. II. A, when the incident state in
channel $\alpha$ is near resonant to the bound states $|\Phi_{b}\rangle$
and $|\Phi_{b'}\rangle$ in channels $\eta$ and $\eta'$, the Green's
operator $g^{(\alpha\beta)}(E_{\alpha})$ can be approximated as
\begin{eqnarray}
g^{(\alpha\beta)}(E_{\alpha}) & = & \frac{1}{E_{\alpha}+i0^{+}+h}+\frac{|\Phi_{b}\rangle\langle\Phi_{b}|}{E_{\alpha}-\epsilon_{b}}+\frac{|\Phi_{b'}\rangle\langle\Phi_{b'}|}{E_{\alpha}-\epsilon_{b'}}.\nonumber \\
\label{gg3}
\end{eqnarray}
Under this approximation we can solve Eq. (\ref{lse3-1}) with the
method in Appendix B, and derive the expression of $|\Psi_{\sqrt{E_{\alpha}-E_{\beta}},\beta}^{(\alpha\beta\eta'+)}\rangle$:
\begin{eqnarray}
 &  & |\Psi_{\sqrt{E_{\alpha}-E_{\beta}},\beta}^{(\alpha\beta\eta'+)}\rangle=|\Psi_{\sqrt{E_{\alpha}-E_{\beta}},\beta}^{(\alpha\beta+)}\rangle\nonumber \\
 &  & +\frac{g^{(\alpha\beta)}(E_{\alpha})W'|\Phi_{b'}\rangle\langle\Phi_{b'}|W'|\Psi_{k_{\beta},\beta}^{({\rm bg}+)}\rangle}{E_{{\rm s}}-\epsilon_{b'}-\langle\Phi_{b'}|W'g^{(\alpha\beta)}(E_{\alpha})W'|\Phi_{b'}\rangle}.\label{pp2}
\end{eqnarray}
Substituting Eq. (\ref{pp2}) into Eq. (\ref{a3}), we obtain
\begin{equation}
{\bar a}(\epsilon_{b'})=a^{(\alpha\beta\eta')}+\frac{A'}{E_{\alpha}-\epsilon_{b'}-B'},\label{aa3}
\end{equation}
where the parameters $A'$ and $B'$ are given by \begin{widetext}

\begin{eqnarray}
A' & = & \frac{f_{\beta\alpha}^{(\alpha\beta\eta')}(E_{\alpha})\langle\Phi_{b}|W|\Psi_{0,\alpha}^{(\alpha\beta\eta'+)}\rangle\left[E_{{\rm s}}-\epsilon_{b'}-\langle\Phi_{b'}|W'g^{(\alpha\beta)}(E_{\alpha})W'|\Phi_{b'}\rangle\right]}{\langle\Phi_{b}|W|\Psi_{\sqrt{E_{\alpha}-E_{\beta}},\beta}^{(\alpha\beta+)}\rangle},\label{ap}\\
B' & = & \langle\Phi_{b'}|W'g^{(\alpha\beta)}(E_{\alpha})W'|\Phi_{b'}\rangle-\frac{\langle\Phi_{b}|Wg^{(\alpha\beta)}(E_{\alpha})W'|\Phi_{b'}\rangle\langle\Phi_{b'}|W'|\Psi_{\sqrt{E_{\alpha}-E_{\beta}},\beta}^{(\alpha\beta+)}\rangle}{\langle\Phi_{b}|W|\Psi_{\sqrt{E_{\alpha}-E_{\beta}},\beta}^{(\alpha\beta+)}\rangle}.\label{bp}
\end{eqnarray}
\end{widetext} Eq. (\ref{aa3}) is Eq. (\ref{chi-2}) in our main
text.

\subsection{The parameter $B'$}

Now we prove that the parameter $B'$ in Eq. (\ref{chi-2}) is real.
To this end, we first prove two lemmas.

\textbf{Lemma 1}: The complex phase of the wave function $_{R}\langle{\bf r}|\Psi_{\sqrt{E_{\alpha}-E_{\beta}},\beta}^{(\alpha\beta+)}\rangle$
is ${\bf r}$-independent. That is, $_{R}\langle{\bf r}|\Psi_{\sqrt{E_{\alpha}-E_{\beta}},\beta}^{(\alpha\beta+)}\rangle$
can be expressed as $_{R}\langle{\bf r}|\Psi_{\sqrt{E_{\alpha}-E_{\beta}},\beta}^{(\alpha\beta+)}\rangle=|f({\bf r})\rangle_Ie^{i\phi},$
where $|f({\bf r})\rangle_I$ is a real function of ${\bf r}$, and $\phi$ is an ${\bf r}$-independent
constant.

\textit{Proof}: We define an operator $g_{s}^{(\alpha\beta)}(E)$
as
\begin{equation}
g_{s}^{(\alpha\beta)}(E)=P_{s}g^{(\alpha\beta)}(E)P_{s},\label{gs-1}
\end{equation}
where $P_{s}$ is the projection operator to the subspace of $s$-wave
states. It is clear that $g_{s}^{(\alpha\beta)}(E)$ can be re-expressed
as \cite{st}
\begin{eqnarray}
g_{s}^{(\alpha\beta)}(E) & = & \sum_{l=\alpha,\beta}\int d{\bf k}\frac{|\psi_{k,l}^{(\alpha\beta+)}\rangle\langle\psi_{k,l}^{(\alpha\beta+)}|}{E+i0^{+}-k^{2}-E_{l}}+\sum_{q}\frac{|B_{q}\rangle\langle B_{q}|}{E-B_{q}}\nonumber \\
\label{ga}\\
 & = & \sum_{l=\alpha,\beta}\int d{\bf k}\frac{|\psi_{k,l}^{(\alpha\beta-)}\rangle\langle\psi_{k,l}^{(\alpha\beta-)}|}{E+i0^{+}-k^{2}-E_{l}}+\sum_{q}\frac{|B_{q}\rangle\langle B_{q}|}{E-B_{q}}.\nonumber \\
\label{gb}
\end{eqnarray}
Here $|B_{q}\rangle$ is the $q$-th bound state of the system with
$W=W'=0$, and $B_{q}$ is the energy of $|B_{q}\rangle$. Substituting
$E=E_{\alpha}$ into Eqs. (\ref{ga}) and (\ref{gb}) and using
\begin{equation}
\frac{1}{z\pm i0^{+}}={\cal P}\left(\frac{1}{z}\right)\mp i\pi\delta(z),\ z\in{\rm Reals}\label{eq:}
\end{equation}
with ${\cal P}$ the principal value, we obtain
\begin{eqnarray}
 &  & g_{s}^{(\alpha\beta)}(E_{\alpha})-g_{s}^{(\alpha\beta)\dagger}(E_{\alpha})\nonumber \\
 & = & -i4\pi^{2}\sqrt{E_{\alpha}-E_{\beta}}|\Psi_{\sqrt{E_{\alpha}-E_{\beta}},\beta}^{(\alpha\beta+)}\rangle\langle\Psi_{\sqrt{E_{\alpha}-E_{\beta}},\beta}^{(\alpha\beta+)}|\nonumber \\
\label{r1}\\
 & = & -i4\pi^{2}\sqrt{E_{\alpha}-E_{\beta}}|\Psi_{\sqrt{E_{\alpha}-E_{\beta}},\beta}^{(\alpha\beta-)}\rangle\langle\Psi_{\sqrt{E_{\alpha}-E_{\beta}},\beta}^{(\alpha\beta-)}|.\nonumber \\
\label{r2}
\end{eqnarray}
Thus, we have
\begin{eqnarray}
 &  & _{R}\langle{\bf r}|\Psi_{\sqrt{E_{\alpha}-E_{\beta}},\beta}^{(\alpha\beta+)}\rangle\langle\Psi_{\sqrt{E_{\alpha}-E_{\beta}},\beta}^{(\alpha\beta+)}|{\bf r}'\rangle_{R}\nonumber \\
 & = & _{R}\langle{\bf r}|\Psi_{\sqrt{E_{\alpha}-E_{\beta}},\beta}^{(\alpha\beta-)}\rangle\langle\Psi_{\sqrt{E_{\alpha}-E_{\beta}},\beta}^{(\alpha\beta-)}|{\bf r}'\rangle_{R},\ {\rm for}\ \forall{\bf r},{\bf r^\prime}.\nonumber \\
\label{psi2}
\end{eqnarray}
Since
\begin{equation}
\langle\Psi_{\sqrt{E_{\alpha}-E_{\beta}},\beta}^{(\alpha\beta-)}|{\bf r}\rangle_{R}=\langle\Psi_{\sqrt{E_{\alpha}-E_{\beta}},\beta}^{(\alpha\beta+)}|{\bf r}\rangle_{R}^{\ast},\label{psi3}
\end{equation}
the result (\ref{psi2}) implies that the complex phase of $_{R}\langle{\bf r}|\Psi_{\sqrt{E_{\alpha}-E_{\beta}},\beta}^{(\alpha\beta+)}\rangle$
is ${\bf r}$-independent. $\square$

\textbf{Lemma 2}: In our system the function
\begin{eqnarray}
F(E,{\bf r},{\bf r}') & = & \sum_{l=\alpha,\beta}{\cal P}\int d{\bf k}\frac{_{R}\langle{\bf r}|\psi_{k,l}^{(\alpha\beta+)}\rangle\langle\psi_{k,l}^{(\alpha\beta+)}|{\bf r}'\rangle_{R}}{E-k^{2}-E_{l}}\nonumber \\
 &  & +\sum_{q}\frac{_{R}\langle{\bf r}|B_{q}\rangle\langle B_{q}|{\bf r}'\rangle_{R}}{E-B_{q}},\label{ff}
\end{eqnarray}
with $|B_{q}\rangle$ defined below Eq. (\ref{gb}), is real for any
energy $E$ and any positions ${\bf r}$ and ${\bf r}'$.

\textit{Proof}: Since all the potentials $V_{\alpha\beta}(r)$ are
real functions of $r$, the bound-state wave function $_{R}\langle{\bf r}|\phi_{q}\rangle$
can be chosen to be real. Thus, the second term in the r.h.s of Eq.
(\ref{ff}) is real. Furthermore, Eqs. (\ref{ga}) and (\ref{gb})
imply
\begin{eqnarray}
 &  & \sum_{l=\alpha,\beta}{\cal P}\int d{\bf k}\frac{_{R}\langle{\bf r}|\psi_{k,l}^{(\alpha\beta+)}\rangle\langle\psi_{k,l}^{(\alpha\beta+)}|{\bf r}'\rangle_{R}}{E-k^{2}-E_{l}}\nonumber \\
 & = & \sum_{l=\alpha,\beta}{\cal P}\int d{\bf k}\frac{_{R}\langle{\bf r}|\psi_{k,l}^{(\alpha\beta-)}\rangle\langle\psi_{k,l}^{(\alpha\beta-)}|{\bf r}'\rangle_{R}}{E-k^{2}-E_{l}}.\label{pv}
\end{eqnarray}
From this result and Eq. (\ref{psi3}), the first term in the r.h.s
of Eq. (\ref{ff}) is also real. Therefore, $F(E,{\bf r},{\bf r}')$ is
a real function. $\square$

Based on these two lemmas, now we prove that the parameter $B'$ is
real. We first rewrite the expression (\ref{bp}) of $B'$ as
\begin{eqnarray}
 &  & B^\prime=\langle\Phi_{b'}|W'g_{s}^{(\alpha\beta)}(E_{\alpha})W'|\Phi_{b'}\rangle\nonumber \\
 &  & -\frac{\langle\Phi_{b}|Wg_{s}^{(\alpha\beta)}(E_{\alpha})W'|\Phi_{b'}\rangle\langle\Phi_{b'}|W'|\Psi_{\sqrt{E_{\alpha}-E_{\beta}},\beta}^{(\alpha\beta+)}\rangle}{\langle\Phi_{b}|W|\Psi_{\sqrt{E_{\alpha}-E_{\beta}},\beta}^{(\alpha\beta+)}\rangle}.\nonumber \\
\label{rbp}
\end{eqnarray}
Since in our system all the potentials are real, the wave functions
$_{R}\langle{\bf r}|W'|\Phi_{b'}\rangle$ and $_{R}\langle{\bf r}|W|\Phi_{b}\rangle$
can be chosen as real functions. Therefore, our lemma 1 implies that
factor $\langle\Phi_{b'}|W'|\Psi_{\sqrt{E_{\alpha}-E_{\beta}},\beta}^{(\alpha\beta+)}\rangle/\langle\Phi_{b}|W|\Psi_{\sqrt{E_{\alpha}-E_{\beta}},\beta}^{(\alpha\beta+)}\rangle$
is real. Furthermore, with Eq. (\ref{eq:}) we can rewrite $g_{s}^{(\alpha\beta)}(E_{\alpha})$
as
\begin{eqnarray}
g_{s}^{(\alpha\beta)}(E_{\alpha}) & = & \sum_{l=\alpha,\beta}{\cal P}\int d{\bf k}\frac{|\psi_{k,l}^{(\alpha\beta+)}\rangle\langle\psi_{k,l}^{(\alpha\beta+)}|}{E_{\alpha}-k^{2}-E_{l}}+\sum_{q}\frac{|\phi_{q}\rangle\langle\phi_{q}|}{E_{\alpha}-B_{q}}\nonumber \\
 &  & -i2\pi^{2}\sqrt{E_{\alpha}-E_{\beta}}|\Psi_{\sqrt{E_{\alpha}-E_{\beta}},\beta}^{(\alpha\beta+)}\rangle\langle\Psi_{\sqrt{E_{\alpha}-E_{\beta}},\beta}^{(\alpha\beta+)}|.\nonumber \\
\label{gg}
\end{eqnarray}
Substituting Eq. (\ref{gg}) into Eq. (\ref{rbp}) and using lemma
2, we can directly obtain ${\rm Im}[B']=0$.


\end{subappendices}


\begin{thebibliography}{10}
\bibitem{fano2} According to the 1st-order perturbation theory, the
inelastic scattering amplitude is proportional to the transition matrix
element $\int dX\psi_{f}^{\ast}(X)V_{c}(X)\psi_{i}(X)$, where $X$
is the coordinate of that scattering problem, $V_{{\rm c}}(X)$ is
the IC-OC coupling, $\psi_{i}(X)$ is the incident wave function,
and $\psi_{f}(X)$ is the component of the energy-conserved output
wave function in the OC. In Ref. \cite{fano}, $\psi_{f}(X)$ and
$\psi_{i}(X)$ are denoted as $|\Psi_{E})$ and $|i)$, respectively.
In Sec. II of that reference, Ugo Fano calculated the modification
of $\psi_{f}(X)$ and the transition matrix element by the coupling
to the isolated bound state.
\end{thebibliography}
\end{document}